\documentclass[pra,aps,twocolumn,superscriptaddress,amsfonts,citeautoscript,nofootinbib,a4paper]{revtex4}

\usepackage{amsmath}  \usepackage{amssymb}  \usepackage{amsfonts}  \usepackage{bm}  \usepackage{bbm}   \usepackage{braket}  \usepackage{color}  \usepackage{comment}  \usepackage{dcolumn}  \usepackage{enumerate}  \usepackage{epsfig}  \usepackage{gensymb}  \usepackage{graphicx}  \usepackage{indentfirst}  \usepackage{lmodern}  \usepackage{mathrsfs}  \usepackage{mathtools}  \usepackage{psfrag}  \usepackage{pst-all}  \usepackage{soul}  \usepackage{units}  \usepackage{xcolor}
\usepackage[colorlinks,linkcolor=blue,citecolor=blue,urlcolor=blue,hyperindex,driverfallback=dvipdfm]{hyperref}  \usepackage[T1]{fontenc} 

\def\ii{{\rm i}}  \def\ee{{\rm e}}

\def\rb{{\bf r}}  \def\Rb{{\bf R}}    \def\vb{{\bf v}}
    \def\zz{\hat{\bf z}}

\def\kb{{\bf k}}       
 \def\vb{{\bf v}} 
    \def\gb{{\bf g}}
\def\me{m_{\rm e}}  \def\kB{{k_{\rm B}}}
      \def\Ab{{\bf A}}

\usepackage{lipsum}

\def\tb{\tilde{\beta}}

\begin{document}

\title{Optical-cavity mode squeezing by free electrons}

\author{Valerio~Di~Giulio}
\email{digiuliovalerio@gmail.com}
\affiliation{ICFO-Institut de Ciencies Fotoniques, The Barcelona Institute of Science and Technology, 08860 Castelldefels (Barcelona), Spain}
\author{F.~Javier~Garc\'{\i}a~de~Abajo}
\email{javier.garciadeabajo@nanophotonics.es}
\affiliation{ICFO-Institut de Ciencies Fotoniques, The Barcelona Institute of Science and Technology, 08860 Castelldefels (Barcelona), Spain}
\affiliation{ICREA-Instituci\'o Catalana de Recerca i Estudis Avan\c{c}ats, Passeig Llu\'{\i}s Companys 23, 08010 Barcelona, Spain}

\begin{abstract}
The generation of nonclassical light states bears a paramount importance in quantum optics and is largely relying on the interaction between intense laser pulses and nonlinear media. Recently, electron beams, such as those used in ultrafast electron microscopy to retrieve information from a specimen, have been proposed as a tool to manipulate both bright and dark confined optical excitations, inducing semiclassical states of light that range from coherent to thermal mixtures. Here, we show that the ponderomotive contribution to the electron-cavity interaction, which we argue to be significant for low-energy electrons subject to strongly confined near-fields, can actually create a more general set of optical states, including coherent and squeezed states. The post-interaction electron spectrum further reveals signatures of the nontrivial role played by $A^2$ terms in the light-matter coupling Hamiltonian, particularly when the cavity is previously excited by either chaotic or coherent illumination. Our work introduces a disruptive approach to the creation of nontrivial quantum cavity states for quantum information and optics applications, while it suggests unexplored possibilities for electron beam shaping.
\end{abstract}

\maketitle
\date{\today}

\section{Introduction}

The generation of different states of light is of fundamental interest in quantum optics and enables powerful applications such as the increase in sensitivity achieved in the interferometric detection of gravitational waves through the use of squeezed states with reduced uncertainty \cite{W1983}. Likewise, the generation of approximate Gottesman-Kitaev-Preskill (GKB) states \cite{GKP01} is needed to implement fault-tolerant quantum computing on photonic setups \cite{M14_3}. In this context, quantum two-photon states are commonly produced by exploiting the nonlinear response of some materials to coherent laser illumination through processes such as four-wave mixing \cite{SHY1985} and parametric down-conversion \cite{WKH1986} assisted by bright modes in optical cavities. In a radically different approach, free electrons have been identified as a promising tool to generate quantum light states \cite{paper180,FHA22}. In addition to the nanometer precision with which electron beams (e-beams) can be focused on an optical cavity, they are advantageous with respect to external light excitation in that they can interact with strongly-confined dark modes because they act as broad sources of evanescent fields \cite{paper149}.

The interest in generating low-uncertainty states through e-beams dates back to the early stages in the development of free-electrons lasers, when several theoretical works predicted this kind of statistics in the free-space radiation emission from a linear wiggler \cite{BSZ1982,GB1987,GB91}. More recently, with the advent of ultrafast electron microscopy \cite{LSZ05,BZ15,ARM20}, the experimental ability of integrating intense laser sources and e-beams in a single setup allowed for their synchronized interaction at the specimen, giving birth to the so-called photon-induced near-field electron microscopy (PINEM) \cite{BFZ09}. In this technique, the strong inelastic electron coupling to fields scattered by the specimen results in the absorption and emission of multiple light quanta, causing a substantial reshaping of the electron wave function as well as the state of cavity modes targeted by the laser at the specimen.

From the point of view of the electron, such reshaping consists in the emergence of intense sidebands in the electron spectrum spaced by the laser photon energy \cite{FES15}. Interestingly, upon further propagation over macroscopic distances, coherent electron components possessing different energies (and velocities) evolve into a train of attosecond pulses \cite{PRY17,MB18_2,B17_2}. In addition, our ability to manipulate the electron density matrix can be extended by using squeezed light instead of coherent laser sources \cite{paper360}.

From the perspective of an optical mode in the specimen, the electron is known to simply act as a displacement operator on the coherent state produced by the external illumination \cite{paper339}. In particular, the electron-cavity interaction leads to a density matrix with Poissonian diagonal elements when the cavity is initially prepared in the ground state \cite{paper360}.  Although more complicated states can be obtained from consecutive interactions with multiple electrons combined with projective measurements onto specific electron states \cite{HRN21,DBG22}, single electrons have so far been regarded as a semiclassical current acting on the specimen. This is indeed a robust assumption for energetic electrons whose velocity $\vb$ is negligibly perturbed by the interaction (nonrecoil approximation), provided linear terms in the electron-light coupling Hamiltonian (i.e., $\propto\vb\cdot\Ab$, where $\Ab$ is the vector potential) are dominant over ponderomotive $\propto A^2$ contributions. This condition is however challenged for an electric field associated with a mode polarized normally to the electron propagation direction, or when sign cancellations render a vanishing linear coupling coefficient, as well as when low-energy electrons such as those used in low-energy electron microscopy (LEEM) \cite{R95,NYI06} ($\lesssim 100$ eV) are made to interact with cavities supporting resonances in the infrared range. In fact, when linear and quadratic $\Ab$ terms have similar strengths in the coupling Hamiltonian, we expect the effect of a single electron on a cavity mode to deviate from the classical-current description, leading, for instance, to the direct production of squeezed cavity states.

\begin{figure}
\includegraphics[scale=0.8]{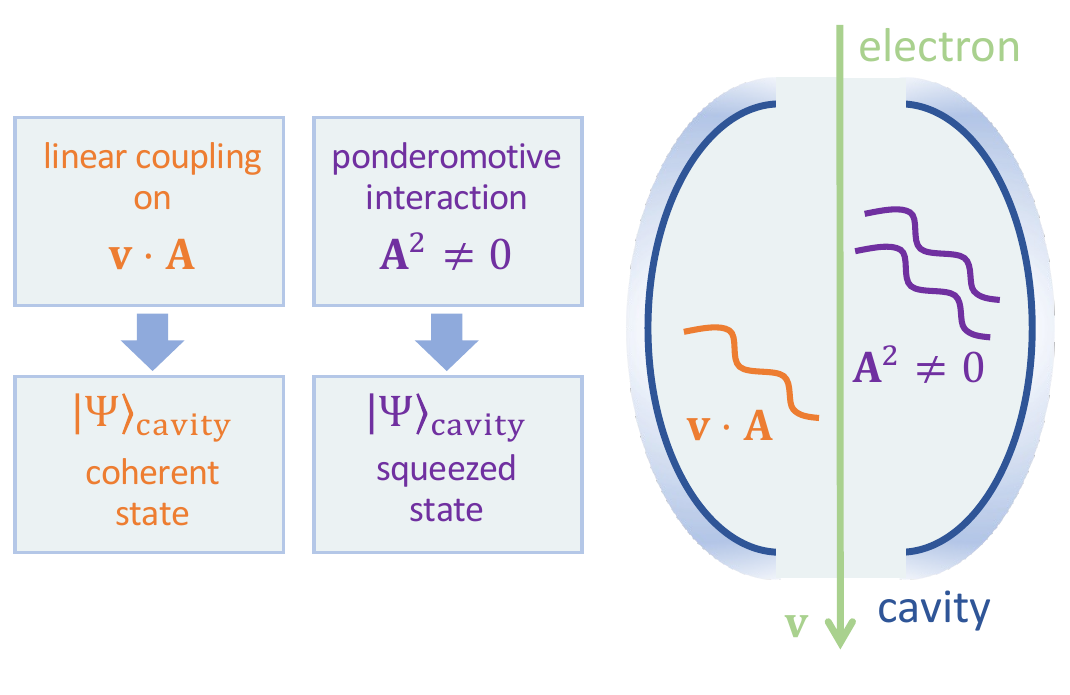}
\caption{{\bf Free-electron interaction with an optical cavity}. We present a sketch of the interaction for a single-mode cavity. Both linear ($\propto\vb\cdot\Ab$) and quadratic (ponderomotive $\propto A^2$) terms in the quantum vector potential operator $\Ab$ are present in the minimal-coupling light-matter interaction Hamiltonian. Switching on and off these two terms selects the creation of either coherent or squeezed cavity mode states, respectively.
}
\label{Fig1}
\end{figure}
Here, we study the interaction of a free electron with a single-mode cavity under conditions in which linear and ponderomotive interactions have commensurate strengths or when the latter dominates (Fig.\ \ref{Fig1}). We start by developing the theory describing the interaction of a fast electron with a single-mode cavity in the nonrecoil approximation including the effect of the quantized ponderomotive interaction. Remarkably, the entire dynamics admits an analytical solution in terms of the consecutive action of a displacement and a squeezing operator. Then, we theoretically demonstrate that a cavity starting from its ground state is left in a general two-photon coherent state \cite{Y1976} after interaction with a single electron, including vacuum-, phase-, and amplitude-squeezed states. Finally, we study how the presence of the ponderomotive interaction can lead to asymmetric distributions in the electron spectrum both when the cavity is initially heated (i.e., prepared in a thermal state) or when it is irradiated by laser light (i.e., in an initial coherent state). Besides their relevance from a fundamental viewpoint, our results suggest a way to produce nonclassical states of light in dark modes that are accessible to electrons by coupling to the evanescent components of the electric field accompanying the charged probe.  

\section{Electron-cavity quantum dynamics}
\label{ECQD}
\subsection{The evolution operator}
Following the methods introduced in Ref. \cite{paper339}, we start by considering a structure characterized by a single photonic mode of energy $\hbar\omega_0$ satisfying a bosonic statistics and interacting with an incoming electron initially prepared with a wave function $\psi_0(\rb,t)=\ee^{\ii (E_0 t/\hbar - \kb_0\cdot \rb)}\phi_0(\rb -\vb t)$. This expression assumes that $\psi_0(\rb,t)$ is made of momentum components that are closely peaked around a value $\hbar \kb_0$ corresponding to a central energy $E_0=c\sqrt{\me^2c^2+\hbar^2 k_0^2}$, such that the function $\phi_0(\rb -\vb t)$ varies slowly in space and time. By further assuming the lifetime of the photonic mode to be much longer than the electron-specimen interaction time, and adopting the nonrecoil approximation, which is valid for keV electrons and cavities resonating at visible or lower frequencies, the dynamics of the combined electron-cavity system can be well described by the Schr\"{o}dinger equation $\ii \hbar \partial_t |\psi(\rb,t)\rangle=(\hat{\mathcal{H}}_0+\hat{\mathcal{H}}_{\rm int})|\psi(\rb,t)\rangle$ with free and interaction Hamiltonians given by (see Appendix\ \ref{secA1})
\begin{subequations}
\begin{align}
\hat{\mathcal{H}}_0 &=\hbar \omega_0 a^\dagger a + E_0 -\hbar \vb \cdot \left(\ii \nabla + \kb_0\right), \label{H0}\\
\hat{\mathcal{H}}_{\rm int}& =(e \vb /c) \cdot \hat{\Ab}(\rb)+ \sum_i g_i \hat{A}^2_i(\rb), \label{Hint}
\end{align}
\end{subequations}
where $\hat{\Ab}(\rb)=(-\ii c/\omega_0)\left[\vec{\mathcal{E}}_0(\rb)a-\vec{\mathcal{E}}_0^*(\rb)a^\dagger\right]$ is the quantized vector potential operator incorporating the normalized electric field mode profile $\vec{\mathcal{E}}_0(\rb)$, the sum over $i$ runs over Cartesian components, and the vector $\gb=(e^2/2\me c^2 \gamma)(1,1,\gamma^{-2})$ with $\gamma=1/\sqrt{1-(v/c)^2}$ introduces approximate relativistic corrections in the ponderomotive interaction.

As explained in detail in Appendix\ \ref{secA2}, given the initial condition $|\psi(\rb,t\rightarrow -\infty)\rangle =\psi_0(\rb,t)\sum_n \alpha^0_n \ee^{-\ii n\omega_0 t}|n\rangle$ (i.e., uncorrelated electron and cavity states), the problem admits an analytical solution that represents a generalization to the one found in Ref.\ \cite{paper339}, including the effect of the single-mode ponderomotive force [the $A^2$ terms in Eq.\ (\ref{Hint})]. More precisely, taking the electron velocity as $\vb = v \zz$, we find 
\begin{align}
|\psi(\rb,t)\rangle=\psi_0(\rb,t)\sum_{\ell ,n}\ee^{\ii \omega_0 [\ell(z/v-t)-nt]}\ee^{\ii \theta_n(\rb)}f_\ell^n(\rb)|n \rangle, \label{solution}
\end{align}
where the phases $\theta_n=-\sum_i \int_{-\infty}^z dz' |\eta_{i}(z')|^2 (2n+1)$ incorporate a dependence on the mode field and electron velocity through
\begin{align}
\eta_i(z)= \sqrt{\frac{g_i c^2}{v\hbar\omega_0^2}}\,\mathcal{E}_{0,i}(z)\ee^{-\ii \omega_0 z/v}.
\label{etai}
\end{align}
For the sake of simplicity, we hereinafter do not explicitly indicate the dependence on the electron position vector $\rb$. The expansion coefficients $f_\ell^n$ in Eq.\ (\ref{solution}) can be written as
\begin{align}
f_\ell^m(\rb)&=\alpha^0_{m+\ell} \,\ee^{\ii \tilde{\chi} } \, F_{\ell}^n \label{fln}
\end{align}
in terms of the matrix elements
\begin{align}
F_{\ell}^n&= \langle m | \hat{S}(\sigma_0)\,\ee^{-\ii \lambda (a^\dagger a+a a^\dagger)/2 }\hat{D}(\tilde{\beta}_0)|m+ \ell \rangle, \label{Fl}
\end{align}
where $\hat{S}(\sigma_0)=\ee^{\sigma_0 \,a a/2-\sigma_0^* \,a^\dagger a^\dagger/2}$ and $\hat{D}(\tilde{\beta}_0)=\ee^{\tilde{\beta}_0^*a^\dagger -\tilde{\beta}_0 a }$ are squeezing and displacement operators, respectively. Incidentally, an overall phase $\tilde{\chi}$ emerges in Eq.\ (\ref{fln}) as a result of the retarded elastic image interaction between the electron and the cavity \cite{paper357}, but such phase is irrelevant for this study. In Eq.\ (\ref{Fl}), we introduce the squeezing and displacement coefficients
\begin{subequations}
\label{slbcoeff}
\begin{align}
\sigma_0&={\rm arcsinh}(|\nu|)\,\ee^{\ii(\arg\{\mu\}-\arg\{\nu\})}, \\
\tilde{\beta}_0&=\int_{-\infty}^z dz' \Big[\ee^{2\ii\theta_{0}(z')}\nu^*(z')\partial_{z'}\beta^*_0(z') \\
&\quad\quad\quad\quad\quad+\ee^{ -2 \ii \theta_{0}(z')}\mu^*(z')\partial_{z'} \beta_0(z')\Big], \nonumber 
\end{align}
\end{subequations}
as well as $\lambda =\arg\{\mu\}$. These quantities are directly obtained from the solution of the Ricatti differential equation \cite{R1972} $\partial_z R/2=  k^*-k R^2 $ with $R=\nu/\mu$,
\begin{subequations}
\label{munu}
\begin{align}
\mu &=\ee^{2 \int_{-\infty}^z dz' k(z')R(z')},\\
\nu &=2 \int_{-\infty}^z dz' k^*(z') \mu(z'),
\end{align}
\end{subequations}
and $k=\ii \sum_i \eta_i^2 \ee^{4\ii \theta_0}$. Interestingly, by differentiating Eqs.\ (\ref{munu}), we readily verify that $\mu$ and $\nu$ satisfy the unitary condition of a generalized Bogoliubov transformation, $|\mu|^2-|\nu|^2=1$ \cite{BV1986}. Also, when the ponderomotive force is neglected, the displacement factor $\tilde{\beta}_0$ becomes identical with the coupling factor found in Ref. \cite{paper339}: $\beta_0=(e/\hbar\omega_0)\int_{-\infty}^z dz' \,\mathcal{E}_{0,z}(z')\,\ee^{-\ii \omega_0 z'/v}$.

\subsection{Phase-matched interaction}
\label{PMI}

\begin{figure}
\includegraphics[scale=0.92]{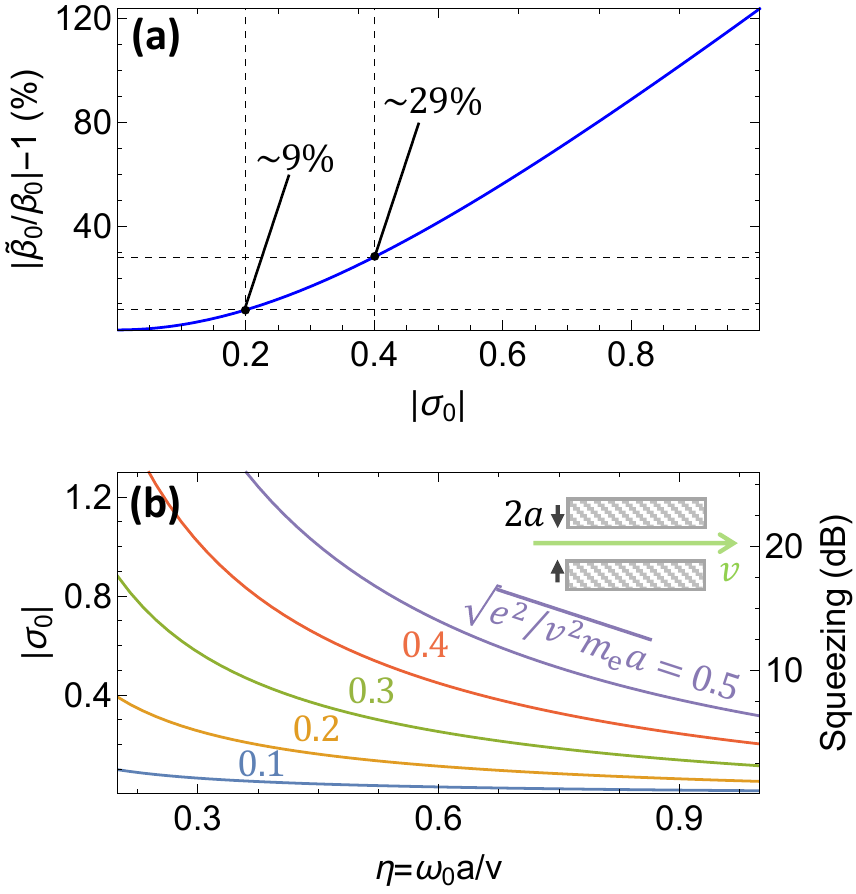}
\caption{{\bf Linear and quadratic coupling in phase-matched interactions}. (a) Deviation of the linear coupling coefficient $|\tb_0|$ from the value obtained by neglecting the ponderomotive force $|\beta_0|$ as a function of $|\sigma_0|$ for a phase-matched interaction with real $\eta_i$ [see Eqs. (\ref{etai}) and (\ref{exactcoupl1})].  (b) Squeezing parameter $|\sigma_0|$ (left vertical axis) and squeezing factor $20|\sigma_0|$\,dB (right vertical axis) as a function of $\eta=\omega_0 a/v$ for different values of $v^2a$ under the electron-cavity configuration depicted in the inset, in which the e-beam moves with velocity $v$ along the axis of a cylindrical hole of radius $a$ and the cavity mode consists of a polariton made of a combination of azimuthal numbers $m=\pm 1$ and phase-matched axial wave vector $q_z=\omega_0/v$.}
\label{Fig2}
\end{figure}

Although the solution in Eq. (\ref{fln}) can be applied to any field profile, here we focus on a simple configuration that allows us to easily understand the problem. Specifically, we consider the electron to move along a path of length $L$ while interacting with an electric field of the form $\vec{\mathcal{E}}_0(\rb)=\vec{\mathcal{E}}_0(\Rb)\ee^{\ii q_z z }$, such that the phase velocity matches the electron velocity (i.e., $q_z=\omega_0/v$). In this scenario, which requires a specimen that possesses translational invariance along the e-beam direction, the ponderomotive coupling factor becomes independent of the longitudinal position $z$, so we have
\begin{align}
k= \ii \sum_i \eta_i^2 \nonumber
\end{align}
[see Eq.\ (\ref{etai}) for $\eta_i$] by disregarding the small phase $\theta_0$. Then, the Ricatti equation admits an analytical solution that directly leads to the following compact forms of the coupling coefficients (see Appendix\ \ref{secB}):
\begin{subequations}
\label{exactcoupl}
\begin{align}
\tb_0 &= \beta_0\;(1+2 \ii L\sum_i |\eta_i|^2)+  \beta_0^* \sigma_0,  \label{exactcoupl1}\\
\sigma_0 &= 2 k L \label{exactcoupl2},
\end{align}
\end{subequations}
where $\beta_0=(eL/\hbar \omega_0)\mathcal{E}_{0,z}(\Rb)$ and $\lambda=0$. Equation\ (\ref{exactcoupl1}) represents a generalization of the intrinsic PINEM coupling coefficient, now also incorporating the effect of the ponderomotive interaction. Interestingly, the presence of $\sigma_0$ can play a significant role in the determination of the linear coupling and, in turn, in the computation of the related EELS probability given by $|\tb_0|^2$, especially for long interaction lengths $L$. A quantitative estimate of such correction can be easily obtained for the special case in which all coefficients $\eta_i$ are real, such that $|\tb_0/\beta_0|=\big|1+2\ii |\sigma_0|\big|$. In Fig. \ref{Fig2}a, we show that the deviation reaches $\sim 9$\% for $|\sigma_0 |\sim 0.2$. Besides the modification of the linear coupling, $\sigma_0$ in Eq.\ (\ref{exactcoupl1}) reflects the magnitude of the effect produced by the squeezing operator [see Eq.\ (\ref{Fl})] on the electron-cavity dynamics.

A key point in this discussion refers to the normalization of the mode field $\vec{\mathcal{E}}_0$, which is determined by the integral $\int d^3\rb\, \epsilon (\rb)|\vec{\mathcal{E}}_0(\rb)|^2=2\pi\hbar \omega_0$ for dielectric cavities made of materials with real permittivity $\epsilon(\rb)$ \cite{GL91}. For instance, for an electric field of real amplitude $E_0$ uniformly distributed in vacuum over an area $A$ and oriented perpendicularly with respect to the e-beam velocity, the normalization condition leads to $E_0=\sqrt{2\pi \hbar \omega_0/LA}$, which in turn yields a squeezing factor $|\sigma_0|=e^2\pi / \omega_0 \me v A$. Unfortunately, we immediately notice that the ponderomotive coupling coefficient is not proportional to the path length $L$, in contrast to the linear PINEM term $\tb_0$ arising when the field has a component along the $z$ direction --a mechanism that has been suggested to enable strong electron-cavity coupling and, from here, electron-mediated entanglement between different cavities \cite{K19}. Nonetheless, the dependence of $\sigma_0$ on the inverse of the electron velocity, the inverse mode frequency, and the field confinement (i.e., $\propto1/A$) should allow us to reach sizeable ponderomotive couplings by going to slow electrons interacting with infrared fields. Indeed, for a mode of energy $\hbar \omega_0=15$\,meV distributed over an area $A=100$\,nm$^2$, a 50\,eV electron produces a squeezing parameter $|\sigma_0|\sim 0.1$.

As a potentially practical scenario, we consider the electron to be focused at the center of a circular hole ($\Rb=0$) of radius $a$ drilled in a polaritonic material (see inset in Fig.\ \ref{Fig2}b) under the aforementioned phase-matching condition. In particular, for a mode consisting of a linear combination of two degenerate modes with azimuthal numbers $m=\pm1$ and a relative phase $\psi$, the linear coupling vanishes (i.e., $\mathcal{E}_{0z}=0$), while the ponderomotive coupling coefficient becomes $\sigma_0=\ii\, \ee^{\ii \psi} e^2/2\me\gamma  v^2 a \eta I(\eta)$ with $\eta =\omega_0 a /v$ and $I(\eta)$ standing for a normalization integral (see Appendix\ \ref{secB1} for details). This dependence clearly favors small radii (i.e., strong field confinement),  small electron velocities, and small resonance energies (see Fig.\ \ref{Fig2}b), in agreement with the scaling properties presented above.

\begin{figure*}
\includegraphics[scale=0.86]{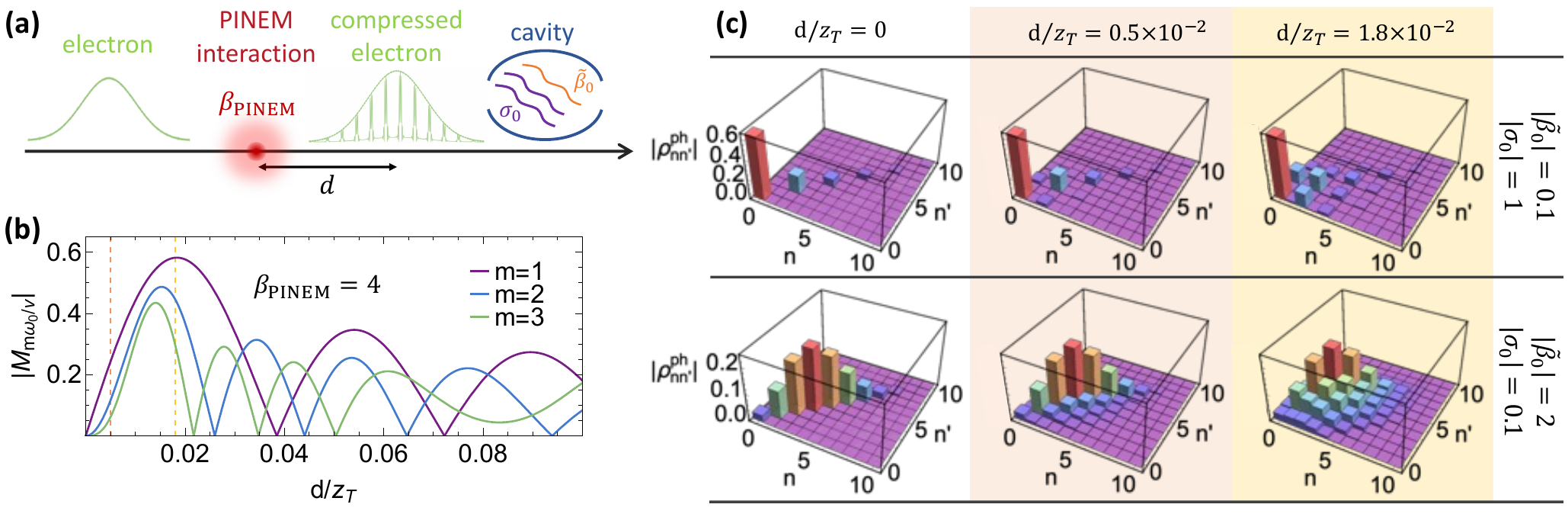}
\caption{{\bf Post-interaction cavity-mode population}. (a) Sketch of an electron wave packet undergoing a classical PINEM interaction of strength $\beta_{\rm PINEM}$, and subsequently propagating in free-space for a distance $d$, where it develops a sequence of probability-density pulses and interacts with a cavity with coupling coefficients $\sigma_0$ and $\tb_0$. (b) $|M_{m \omega_0 /v}|$ factor, determining the amount of coherence left in the two-photon coherent state [see Eq. (\ref{phstate})] for the scenario depicted in panel (a) with $\beta_{\rm PINEM}=4$. We plot the result as a function of the normalized propagation distance $d/z_{\rm T}$ for several harmonics $m$. (c) Elements $\rho_{nn'}^{\rm ph}$ of the density matrix associated with the cavity state after interaction with a compressed electron [see panel (a)]. The interaction region is taken to be placed at the propagation distances highlighted by the colored vertical dashed lines in panel (b) assuming pairs of coupling parameters $|\tb_0|=0.1$ and $|\sigma_0|=1$ (top row); or $|\tb_0|=2$, $|\sigma_0|=0.1$ (bottom row). All plots in panel (c) are calculated for $\phi=0$.}
\label{Fig3}
\end{figure*}

\section{Post-interaction quantum cavity state}

\subsection{Cavity prepared in the ground state}

We now discuss the photonic state after interaction with the electron for the cavity mode starting in the ground state (i.e., $\alpha^0_n=\delta_{n,0}$) and the e-beam being well-focused around a transverse position $\Rb_0$, so that we can approximate $|\psi_0(\rb,t)|^2 \approx \delta(\Rb-\Rb_0)|\phi_0(z-vt)|^2$. In addition, because we are interested in the state of the system after the electron has abandoned the interaction region, we let the longitudinal coordinate $z$ go to infinity in the calculation of all coupling parameters. 

By starting from Eq. (\ref{solution}), we first obtain the full density matrix $\hat{\rho}(\rb,\rb ',t)=|\psi(\rb,t)\rangle \langle \psi(\rb ',t)|$, and then, by integrating over the electron coordinates, we find that the density matrix of the cavity mode reduces to $\int d^3\rb \,\hat{\rho}(\rb,\rb,t)=\sum_{nn'}\ee^{\ii (n'-n) \omega_0 t } \rho_{nn'}^{\rm ph}|n\rangle \langle n' |$, with matrix elements $\rho_{nn'}^{\rm ph}=M_{\omega_0(n'-n)/v}\,\ee^{\ii (\theta_n-\theta_{n'})} F^{n}_{-n} F^{n' *}_{-n'} $ that involve the coherence factor $M_{\omega/v}=\int_{-\infty}^\infty dz \, |\phi_0(z)|^2\, \ee^{\ii \omega z /v}$. The latter, which only depends on the e-beam probability density, determines the ability of the system to interfere with light synchronized with the electron pulse \cite{paper373,paper374}. We remark that this procedure renders the correct statistical properties of the cavity when no measurement is performed on the electron \cite{NC04}.

Interestingly, the coefficients $F^n_{-n}$ turn out to be the projections of the well-known two-photon coherent state defined in a seminal paper by Yuen \cite{Y1976}, from which we find
\begin{subequations}
\label{phstateall}
\begin{align}
\rho_{nn'}^{\rm ph}&= \mathcal{N}^2\,M_{\omega_0(n'-n)/v}\,\ee^{\ii \Delta_{nn'}}  \label{phstate} \\
&\times (n! n'!)^{-1/2} \left({\rm tanh}|\sigma_0|/2\right)^{(n+n')/2}   H_n(\zeta) H^*_{n'}(\zeta), \nonumber
\end{align}
where $H_n$ are Hermite polynomials and we have defined
\begin{align}
\Delta_{nn'}&=\theta_n-\theta_{n'} +(n'-n)\,{\rm arg}\{\sigma_0\}/2, \\
\zeta&=|\tb_0|\ee^{\ii \phi}/\sqrt{\sinh(2|\sigma_0|)}, \\
\phi&={\rm arg}\{\sigma_0\}/2-{\rm arg}\{\tb_0\}-\lambda, \label{phaseshift}
\end{align}
as well as the normalization constant
\begin{align}
\mathcal{N}^2&=\ee^{-|\tb_0|^2[1-\cos (2\phi)\tanh |\sigma_0|]}/\cosh |\sigma_0|.
\end{align}
\end{subequations}
Equation (\ref{phstate}) tells us that different post-interaction cavity states can be selected by properly tuning the factors $\tb_0$ and $k$. For instance, in a geometry for which $\vec{\mathcal{E}}_0\cdot \vb \sim 0$, as in the case analyzed in the previous section, where an electron traverses a circular hole, the coupling coefficient $\tilde{\beta}_0$ vanishes and the state assumes the form of a squeezed vacuum containing only an even number of photons in the diagonal of the density matrix. This conclusion can be easily verified by evaluating the Hermite polynomials at zero argument or, alternatively, directly from Eq.\ (\ref{fln}). In contrast, the off-diagonal elements bear a more intricate dependence on the incoming electron wave function given by the coherence factor $M_{\omega_0 m /v}$, which determines the possibility of creating squeezed states with the correct coherences. For instance, an electron that has undergone a PINEM interaction with an optical field of photon energy $\hbar \omega_0$ and coupling strength $|\beta_{\rm PINEM}|$, followed by free propagation along a macroscopic distance $d$ (see Fig.~\ref{Fig3}a), has an associated probability density consisting of a train of attosecond pulses yielding $|M_{\omega_0 m /v}|=|J_m [4 |\beta_{\rm PINEM}|\sin ( 2\pi m d/z_{\rm T})]|$, where $z_{\rm T}=4\pi \me v^3 \gamma^3/\hbar \omega_0$ is the so-called Talbot distance \cite{ZSF21,paper373}. We show the resulting coherence factor in Fig.~\ref{Fig3}b. The corresponding state left in the cavity after interacting with the electron (top row of Fig.~\ref{Fig3}c) comprises off-diagonal elements that become visible only when $|M_{\omega_0 m/v}|$ takes values around its maximum along the propagation distance ($|M_{\omega_0 m/v}|\sim 0.58$ for $m=1$). Despite this limitation, values of $|M_{\omega_0 m/v}|\sim 1$ have been shown to be realizable by means of a sequence of several cycles of PINEM interaction and free-space propagation \cite{YFR21}. From the expression of $\sigma_0$ related to such interaction, and by using the state in Eq. (\ref{phstate})  with full electron coherence ($M_{\omega_0 m /v}=1$) as well as by setting $\tb_0= 0$, $\lambda=0$ and $\theta_n=0$, we compute the variance $\langle \Delta \hat{X}^2 \rangle =\langle \hat{X}^2\rangle-\langle \hat{X}\rangle^2$ with $\hat{X}=(a+a^\dagger)/2$ --a quantity of utmost importance, which represents the fluctuations in the position of the harmonic oscillator. By choosing $\psi=-\pi/2$ to make $\sigma_0$ real, we obtain $\langle \Delta \hat{X}^2 \rangle = \ee^{-2 |\sigma_0|}/4$. In Fig. (\ref{Fig2}b), we present the degree of squeezing in decibel (vertical right axis) through the relation $-10 \log \big(4 \langle \Delta \hat{X}^2 \rangle \big)=20 |\sigma_0| $. Interestingly, the squeezing approaches values as high as $\sim 10$ dB for a $5$ eV electron, $a\sim 3$ nm, and a resonance energy $\hbar \omega_0 \sim 60$ meV; this squeezing value represents a threshold of particular importance to grant us access into fault-tolerant quantum algorithms by means of GKB codes \cite{TBM20}.

\indent In the opposite extreme ($k\rightarrow 0$), from Eqs.\ (\ref{slbcoeff}a) and (\ref{munu}),  we obtain $|\sigma_0|\rightarrow 0$, which leads to a Poissonian distribution in $P_n=\rho^{\rm ph}_{nn}$ by applying the asymptotic form of $H_n(\zeta)$ or, again, by simply taking the same limit in Eq.\ (\ref{fln}). This behavior is particularly evident in the bottom row of Fig.\ \ref{Fig3}c, where we plot the mode population obtained upon direct evaluation of Eq.\ (\ref{phstate}) for different propagation distances from the first PINEM interaction. 

For parameters $\tb_0$ and $\sigma_0$ lying in between the two extremes under consideration, the cavity state varies in a continuous fashion, spanning values of the zero-delay second-order correlation function $g^{\rm(2)}(0)=\big[\sum_nn(n-1)P_n\big]\big/\big(\sum_n nP_n\big)^2$ ranging from a super-Poissonian statistics with $g^{\rm (2)}(0)\sim 3+{\rm sinh}^{-2}(|\sigma_0|)$ to a Poissonian statistics with $g^{\rm (2)}(0)\sim 1$ when the phase shift is $\phi=\pi/2$ [see Eq.\ (\ref{phaseshift})], as well as to a sub-Poissonian statistics for $\phi=0$. This is a consequence of the decreasing variance of the distribution of photon numbers, $\Delta n^2 = 2\sinh^2(|\sigma_0|)\cosh^2(|\sigma_0|) + |\tb_0|^2 [\cosh(4|\sigma_0|)-\cos(2\phi)\sinh(4|\sigma_0|)]$. We remark that the exponential dependence of the variance on the squeezing parameter renders the post-interaction cavity state extremely sensitive to any change in the phase $\phi$, even for small $|\sigma_0|$, provided the interaction is assisted by a high coupling with the electric field component along the e-beam direction.
\subsection{Cavity under laser illumination}
Although the optical cavity mode may in principle be chosen such that it constrains the phase $\phi$ [Eq.\ (\ref{phaseshift})], in practical configurations this parameter would be hard to change in a single experimental setup. A more feasible route to gain control over the phase $\phi$ consists in aligning the e-beam such that $\mathcal{E}_{0,z}=0$ while the cavity is irradiated by a laser. Intuitively, the post-interaction cavity state should take a form that resemblances the one in Eq.\ (\ref{phstate}), but with $\tb_0$ changed to a factor $\alpha_{\rm L}$ proportional to the incident laser amplitude and carrying its phase. More precisely, the probability to leave $n$ photons in the cavity is
\begin{align}
P_n=\sum_{\ell,\ell'}&M_{\omega_0 (\ell'-\ell)/v} F_\ell^n F_{\ell'}^{n *}  \label{Pnextpump}  \\
\times &\langle n+\ell |\hat{D}(\alpha_{\rm L})|0\rangle \langle 0|\hat{D}^\dagger (\alpha_{\rm L})|n+\ell'\rangle. \nonumber
\end{align}   
In contrast to the scenario with no external pumping, the probabilities $P_n$ in Eq.\ (\ref{Pnextpump}) are strongly dependent on the electron coherence factor. Specifically, Eq.\ (\ref{Pnextpump}) reduces to Eq.\ (\ref{phstate}) only if the electron bears full coherence with respect to the laser [i.e., for $M_{\omega_0 (\ell' -\ell)/v}=1$].
\section{Electron spectrum after interaction with an excited cavity}
We now study the electron spectrum resulting from the interaction with a cavity that is exposed to a synchronized light source. We model this scenario by assuming different initial cavity states (i.e., Fock-state amplitudes $\alpha^0_n$) and an electron whose temporal envelope varies negligibly during one optical period (e.g., for a $\sim100$\,fs electron pulse and a mode energy $\hbar\omega_0\sim 1$\,eV, or equivalently, a mode period $2\pi/\omega_0\approx4.1\,$fs). Under these conditions, once the electron has abandoned the interaction region, its state is approximately given by a discrete superposition of plane waves of energies $E_0+\hbar \omega_0 \ell /v$ labeled by the net number of exchanged photons $\ell$, with associated intensities
\begin{align}
P_\ell=\sum_{n=0 }^\infty |f_\ell^n|^2. \label{Pl}
\end{align}   
The resulting electron energy-loss distribution is $\Gamma(\omega)=\sum_{\ell=-\infty}^\infty P_\ell \,\delta(\omega+\ell \omega_0)$, as obtained directly from Eq.\ (\ref{solution}) by setting $\phi_0(\rb-\vb t)=(2 \pi)^{-3/2}$ and recalling that, in the nonrecoil approximation, the energy is only dependent on the momentum component parallel to the electron trajectory. 

\begin{figure}
\includegraphics[scale=0.7]{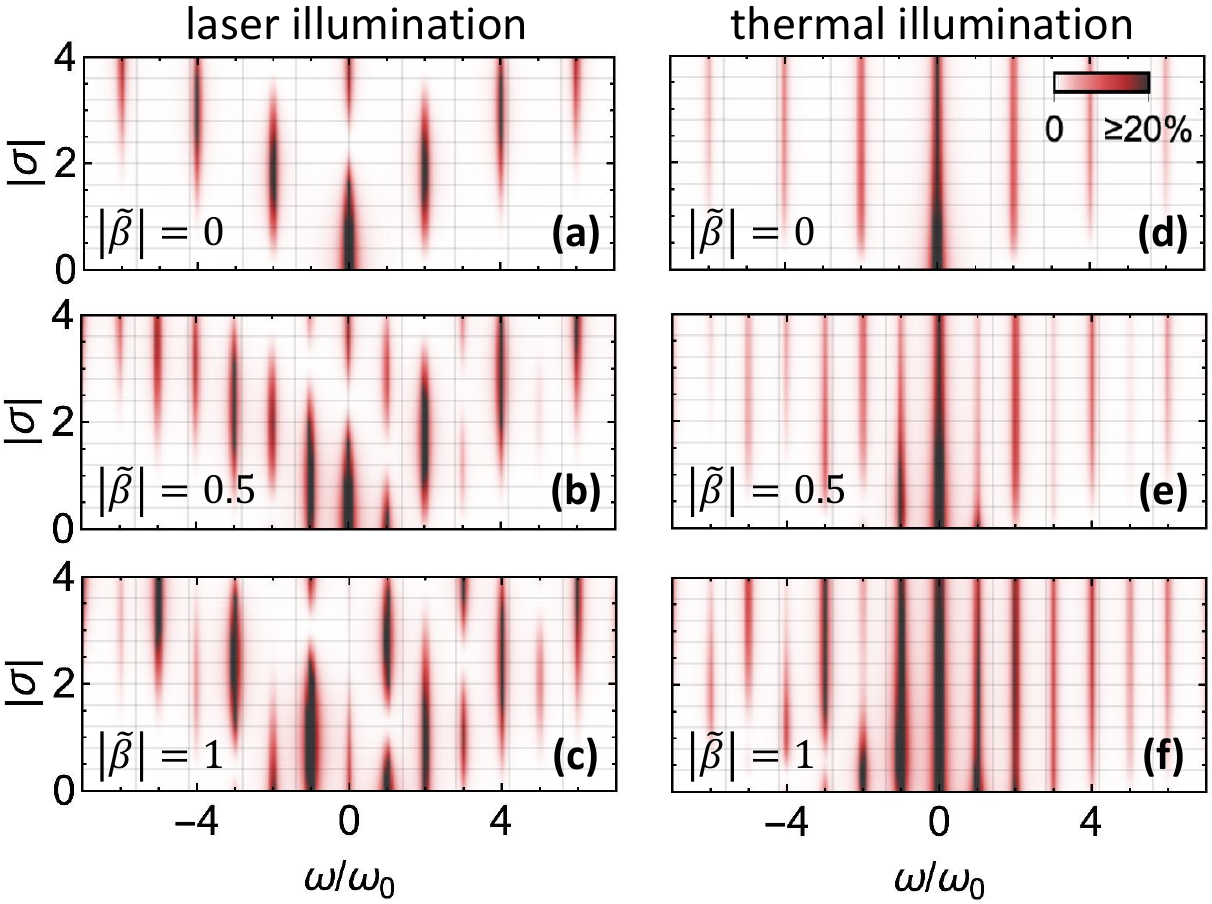}
\caption{{\bf Electron energy-loss spectrum}. Electron distribution as a function of the normalized energy loss $\omega/\omega_0$ and the ponderomotive coupling $|\sigma|$ for $|\tb|=0$ (a,d), $|\tb|=0.5$ (b,e), and $|\tb|=1$ (c,f). We take the cavity to be initially prepared in either a coherent state (a-c) or a thermal state (d-f), as described by Eqs.\ (\ref{Pch}) or (\ref{Pth}), respectively. The conditions $\bar{n}\gg 1$ and $\phi=\pi/2$ are assumed in all panels, and a Lorentzian broadening in $\omega$  (FWHM$=0.12\,\omega_0$) is introduced for clarity.}
\label{Fig4}
\end{figure}

We first consider a quasi-monochromatic laser pulse with central frequency $\omega_0$ irradiating the cavity and inducing an initial coherent state with coefficients $\alpha^0_n = \ee^{-|\alpha_{\rm L}|^2/2}\alpha_{\rm L}^n/\sqrt{n!}$ in the targeted cavity mode. Neglecting any nonlinear response of the cavity materials, we have that $\alpha_{\rm L}$ is proportional to the electric field of the incident laser pulse. In practice \cite{FES15,EFS16}, the light electric field can be as high as $\sim 10^8$\,V/m and produce an average population reaching $\bar{n}=\sum_n n |\alpha^0_n|^2\sim300$ excitations for a mode of energy $\hbar \omega_0=1$\,eV and uniform spatial distribution over a volume of $10^{-3}\,\mu$m$^3$. In this limit, the Poissonian distribution $|\alpha^0_n|^2$ approaches a Gaussian distribution with equal average and standard deviation corresponding to $|\alpha_{\rm L}|^2$, which we further approximate as a Kronecker delta peaked at the integer nearest to $\bar{n}$. Interestingly,  also in the $\bar{n}\gg 1$ limit, and assuming small linear and quadratic coupling coefficients $|\tilde{\beta}_0|$ and $|\sigma_0|$ such that they do not produce large interaction probabilities for $\ell \sim \bar{n}$,  we can write the $F_\ell^n$ coefficients in Eq. (\ref{fln}) as (see Appendix\ \ref{secC})
\begin{align}
F_\ell^n & \approx\,   \, \ee^{ \ii \ell {\rm arg}\{-\tilde{\beta}_0\}-\ii \lambda (n+1/2)} \label{flnlarge} \\
\times& \sum_{n=-\infty}^\infty \ee^{-2\ii n \phi}  J_{-n}\Big[(n+\ell)|\sigma_0|\Big] J_{2n+\ell}\Big[2 \sqrt{n+\ell}\,|\,\tilde{\beta}_0|\Big]. \nonumber
\end{align}
Reassuringly, Eq. (\ref{flnlarge}) reproduces the result obtained for an electron interacting with two mutually coherent fields of frequencies $\omega_0$ and $2\omega_0$ \cite{PRY17,paper347}. Then, plugging this expression into Eq.\ (\ref{Pl}), we obtain the electron probabilities
\begin{align}
P_\ell^{\rm coh}=\sum_{n,n'=-\infty}^\infty \ee^{\ii 2(n'-n) \phi}\,I_{n n' \ell}(1),\label{Pch}
\end{align}
where  $I_{nn' \ell}(x)=J_{-n}\big(x |\sigma|\big)$ $J_{2n+\ell}\big(2\sqrt{x}|\tb|\big)$ $J_{-n'}\big(x|\sigma|\big)$ $J_{2n'+\ell}\big(2\sqrt{x}|\tb|\big)$, and we introduce the factors $|\tb|= \sqrt{\bar{n}}|\tb_0|$ and $|\sigma| =\bar{n}|\sigma_0| $ connecting the one-mode classical coupling parameters to their quantum counterparts $|\tb_0|$ and $|\sigma_0|$.

In Fig. \ref{Fig4}a-c, we show electron spectra calculated from Eq.\ (\ref{Pch}) as a function of $\sigma$ and the normalized energy loss $\omega/\omega_0$ for selected values of $\tb$. For vanishing linear coupling ($\tb=0$), we observe the emergence of several energy sidebands separated by multiples of $2\hbar\omega_0$, as expected from the quadratic interaction with the field. Like in PINEM, the sidebands are symmetrically distributed around the zero-loss peak. In addition, as $|\sigma|$ increases, we observe oscillations in the intensities, also similar to those in PINEM \cite{FES15}, and equally resulting from the interaction between the electron and the coherently excited mode. Once the linear interaction is turned on ($\tb>0$), the interference with quadratic-coupling channels manifests in an asymmetric peak distribution, which is already discernible for small $|\sigma|$ values and can be controlled by tuning the relative phase $\phi$.

Similarly, a highly populated mode can be also obtained via cavity heating at temperatures such that $\kB T /\hbar \omega_0 \gg 1$. In this regime, the cavity is prepared in a pure mixture with a classical Boltzmann probability distribution, which is well approximated by the expression  $p_{n_0}\approx \ee^{-n_0/\bar{n}}/\bar{n}$ with $\bar{n}\approx \kB T/\hbar \omega_0$. We calculate the probability of exchanging an energy $\ell\hbar\omega_0$ by setting $\alpha^0_{n+\ell}=\delta_{n+\ell,n_0}$ in Eq.\ (\ref{flnlarge}) and then summing over all the possible initial number of photons $n_0$, weighted by their respective populations $p_{n_0}$. Since $\bar{n}\gg1$, we approximate this sum to an integral taken over the large extension of the distribution (see Ref.\ \cite{paper339}). Thus, for a thermally excited cavity mode interacting with the electron, we obtain
\begin{align}
P^{\rm th}_\ell=\sum_{n,n'=-\infty}^\infty \ee^{\ii 2(n'-n) \phi}\,\int_0^\infty dx\, \ee^{-x} \,I_{n n' \ell}(x).\label{Pth}
\end{align}
To compare this expression with the results obtained under coherent laser illumination, we evaluate Eq.\ (\ref{Pth}) and plot Fig.\ \ref{Fig4}d-f. As one could infer directly from the form of Eq.\ (\ref{Pth}), the effect of the thermal population is to smear the electron spectra obtained under coherent illumination. As a consequence, the intensity oscillations observed in Fig.\ \ref{Fig4}a-c with increasing $|\sigma|$ are absent in Fig.\ \ref{Fig4}d-f and just replaced by a monotonic decrease with $|\sigma|$.        
\section{Conclusions}
In brief, the inclusion of $A^2$ terms in the interaction between e-beams and optical cavity modes causes a departure from the role of electrons as semiclassical excitation sources. In the absence of such terms, free electrons can only create coherent states by interaction with bosonic optical modes \cite{paper360}, unless a post-selection of the electron state is performed \cite{paper180}. In contrast, the presence of a ponderomotive coupling component embodied in the $A^2$ terms gives rise to a new set of cavity states that include vacuum-, phase-, and amplitude-squeezed states, directly created by interaction with the electron without the need of any post-selection. We also demonstrate that a squeezing of $\sim10$ dB, which represents the current fundamental limit for the implementation of quantum algorithms by means of nonclassical states of light, can be produced in realistic designs of nanostructured polaritonic materials.

In addition, our work shows that the ponderomotive terms allow us to estimate the linear coupling strength $\tb_0$, especially when this parameter reaches large values as a consequence of strong electron-cavity coupling, a condition that has been achieved in recent experiments by phase-matching the electron-cavity mode interaction \cite{KLS20,DNS20,MHK21} and is being theoretically investigated to be implemented in integrated photonic designs \cite{HEK22}.

We understand that the present work paves the way towards the generation of nonclassical states of light with nanometric precision, even in modes that do not couple to light, thus revealing a new tuning knob at the intersection of quantum information and e-beam technologies.

\appendix
\section*{Appendix}

\section{Theoretical description of the quantum electron-cavity interaction in the nonrecoil approximation}
\label{secA}
\renewcommand{\theequation}{A\arabic{equation}}
\renewcommand{\thesubsection}{\arabic{subsection}}

\subsection{The effective Hamiltonian}
\label{secA1}

We start by considering a fast electron interacting with a classical electromagnetic field described in the temporal gauge (i.e., with zero scalar potential) in the presence of a material structure. In our analysis, the initial electron wave function is assumed to be in a superposition of momenta concentrated around a value $\hbar \kb_0$, which corresponds to a relativistic energy $E_0$ and velocity $\vb=\hbar \kb_0 c^2/E_0$. Under these conditions, and for optical excitations in the infrared-visible range, the system dynamics can be modeled by means of the effective Schr\"{o}dinger equation $\ii \hbar \partial_t \psi(\rb,t)=\mathcal{H}(t) \psi(\rb,t)$ with the minimal-coupling Hamiltonian \cite{paper368} 
\begin{align}
\mathcal{H}(t) =& E_0 -\hbar \vb \cdot \left(\ii \nabla + \kb_0\right) \nonumber\\
&+(e \vb /c) \cdot \Ab(\rb,t)+ \sum_{i=x,y,z} g_i A^2_i(\rb,t), \nonumber
\end{align}
where $\Ab(\rb,t)$ is the time-dependent vector potential associated with the external illumination in the presence of the cavity. The coefficient vector $\gb=(e^2/2\me c^2 \gamma)(1,1,\gamma^{-2})$ in the ponderomotive term approximately incorporates relativistic corrections through the Lorentz factor $\gamma=1/\sqrt{1-v^2/c^2}$. The electron motion is then described by the scalar wave function $\psi(\rb,t)$.

In this work, we are interested in a quantum description of the cavity and, thus, adopt a quantum-optics approach \cite{GL91} with the vector potential treated as an operator: $\hat{\bf A}(\rb)=\sum_j (-\ii c/\omega_j)\left[\vec{\mathcal{E}}_{j}(\rb)a_j-\vec{\mathcal{E}}^*_{j}(\rb)a_j^\dagger\right]$, where $j$ runs over cavity modes, $\vec{\mathcal{E}}_j(\rb)$ are the corresponding normalized electric field distributions, and we introduce bosonic annihilation and creation operators $a_j$ and $a_j^\dagger$. By further assuming the sample to undergo negligible inelastic losses, and focusing on the interaction with a spectrally dominant and well-defined mode (labelled by $j=0$, and for simplicity we dismiss the $j$ label in the ladder operators), the electron-cavity system can be described by the modified Schr\"{o}dinger equation $\ii \hbar \partial_t |\psi(\rb,t)\rangle = (\hat{\mathcal{H}}_0+\hat{\mathcal{H}}_{\rm int})|\psi(\rb,t)\rangle $ with
\begin{subequations}
\begin{align}
\hat{\mathcal{H}}_0 =&\hbar \omega_0 a^\dagger a + E_0 -\hbar \vb \cdot \left(\ii \nabla + \kb_0\right),\\
\hat{\mathcal{H}}_{\rm int}=&(\ii e \vb/\omega_0)  \cdot \left[\vec{\mathcal{E}}^*_{0}(\rb)a^\dagger-\vec{\mathcal{E}}_{0}(\rb)a\right] \label{Hint}\\
& -  \sum_i (c^2 g_i/\omega_0^2) \left[\mathcal{E}_{0,i}(\rb)a-\mathcal{E}^*_{0,i}(\rb)a^\dagger\right]^2, \nonumber
\end{align}
\end{subequations}
where we have added the noninteracting Hamiltonian of the cavity mode $\hbar\omega_0 a^\dagger a$ and included the photonic degrees of freedom in the combined state $|\psi (\rb,t)\rangle =\sum_{n=0}^\infty \psi_n(\rb,t) \ee^{-\ii \omega_0 n t}|n\rangle$.

\subsection{Exact analytical solution}
\label{secA2}

The interaction Hamiltonian in Eq.\ (\ref{Hint}) presents a quadratic term in the mode ladder operators, therefore generating a dynamics that substantially differs from the one addressed in previous works \cite{FES15,K19}, where only the linear coupling with the electromagnetic field was taken into account. Despite this difference, by following similar steps as those used in Ref.\ \cite{paper339}, we find an analytical solution for the combined wave function.

We start by inserting the ansatz $\psi_n(\rb,t)=\psi_0(\rb,t)\sum_{\ell=-\infty}^\infty f_\ell^n(\rb)\ee^{\ii \theta_n}\ee^{\ii \omega_0 \ell(z/v-t)}$ into the Schr\"{o}dinger equation, where $\psi_0(\rb,t)=\ee^{-\ii E_0 t/\hbar + \ii \kb_o \cdot \rb }\phi_0(\rb-\vb t)$ is the initial electron wave function,  $\theta_n=-\sum_i \int_{-\infty}^z dz' |\eta_i|^2 (2n +1) $, $\eta_i = \sqrt{g_i \hbar / 2\me v \gamma}\, \mu_i $, and $\mu_i=(e/\hbar \omega_0) \mathcal{E}_{0,i}\ee^{-\ii \omega_0 z /v}$. This leads to the set of one-dimensional differential equations
\begin{align}
\partial_z f_\ell^n =& p^* \sqrt{n}f_{\ell+1}^{n-1} -k^* \sqrt{n(n-1)}f_{\ell+2}^{n-2} \label{diffz}\\
&-p \sqrt{n+1} f_{\ell-1}^{n+1} + k \sqrt{(n+1)(n+2)} f_{\ell-2}^{n+2},\nonumber
\end{align} 
with $p = \mu_z \ee^{2\ii \theta_0}$ and $k=\ii \sum_i \eta_i^2 \ee^{4\ii \theta_0}$.

Importantly, Eq.\ (\ref{diffz}) conserves the total number of excitations, and thus, the set of coefficients defined by $\ell+n=s$ for each choice of an integer $s$ evolves separately. This property allows us to map our problem onto a quantum harmonic oscillator (HO) under nonlinear coupling. Indeed, given the Hamiltonian $\hat{\mathcal{H}}_{\rm HO}=\hat{\mathcal{H}}^0_{\rm HO}+ g_1(t)\, a + g_1^*(t)\, a^\dagger + g_2(t)\, a^2 + g_2^*(t)\, a^{\dagger 2} $, with $\hat{\mathcal{H}}^0_{\rm HO}=\hbar \omega_0 a^\dagger a $, and plugging the general state $|\psi(t)\rangle = \sum_n c_n(t)\ee^{-\ii n\omega_0 t} |n \rangle $ into the associated Schr\"{o}dinger equation, we obtain
\begin{align}
&\ii \hbar \partial_t c_n = s\sqrt{n}c_{n-1} g_1^*\ee^{\ii \omega_0 t}+\sqrt{n (n-1)}c_{n-2}g_2^*\ee^{2\ii \omega_0 t} \label{difft}\\
&+ \sqrt{n+1}c_{n+1}g_1\ee^{-\ii \omega_0 t}+ \sqrt{(n+2) (n+1)}c_{n+2}g_2\ee^{-2\ii \omega_0 t} \nonumber,
\end{align}
which connects to Eq.\ (\ref{diffz}) by means of the transformation $c_n \rightarrow f^n_{s-n}$, $t\rightarrow z$, $\ii g_1 \ee^{-\ii \omega_0 t}/\hbar \rightarrow p$ and $- \ii g_2 \ee^{-2 \ii \omega_0 t}/\hbar \rightarrow k$.
The coefficients $c_n(t)$ can also be expressed in terms of the scattering operator $\hat{\mathcal{S}}(t,-\infty)$ as $c_n(t)=\sum_{m=0}^\infty \langle n | \hat{\mathcal{S}}(t,-\infty) | m \rangle c_m(-\infty)$, assuming an initial state of the harmonic oscillator, written here in the interaction picture \cite{S94} as $|\psi(-\infty)\rangle_{\rm I}=\sum_m c_m(-\infty)|m\rangle$. Consequently, we proceed by evaluating the matrix elements of the scattering operator.

To find the scattering operator, we first notice that an analytical solution to the equation $\ii \hbar \partial_t \,\hat{\mathcal{U}}(t,t_0) = \hat{\mathcal{H}}_{\rm HO}\,\hat{\mathcal{U}}(t,t_0)$ was found by Yuen \cite{Y1976} and is given by \cite{BV1986}
\begin{align}
\hat{\mathcal{U}}(t,t_0) =& \ee^{\ii \tilde{\chi}} \,\hat{S}\left(\sigma_0 \ee^{ 2\ii \omega_0 t}\right)\ee^{-\ii \lambda (a^\dagger a + a a^\dagger)/2}\ee^{-\ii \hat{\mathcal{H}}_{\rm HO}^0(t-t_0)/\hbar} \nonumber\\
&\times\hat{D}\big(\tb_0 \ee^{\ii \omega_0 t_0 }\big), \nonumber
\end{align}
where we have defined the operators $\hat{S}(z)=\ee^{z \,a a /2 -z^* a^\dagger a^\dagger/2 }$ and $\hat{D}(z)=\ee^{z^* a^\dagger -za}$, the linear coupling $\tb_0 = (\ii /\hbar) \int_{t_0}^t dt'( g_1(t') \ee^{ \ii \omega_0 t'}-g_1^*(t') \ee^{\ii \omega_0 t'} \nu^* )$, and the quadratic coupling $\sigma_0 = \arcsin(|\nu|)\,\ee^{\ii({\rm arg}\{\mu \} - {\rm arg}\{\nu \} )} $. The coefficients  $\mu =\ee^{\ii \phi}$ and $\nu= (2\ii/\hbar) \int_{t_0}^t dt' g_2^*(t')\ee^{2 \ii \omega_0 t'} \mu(t') $, with $\phi = (-2/\hbar) \int_{t_0}^t dt'\,g_2(t')\ee^{-2 \ii \omega_0 t'}R(t')$, satisfy the relation $|\mu|^2 - |\nu|^2=1$ at every time, while the function $R=\nu/\mu$ obeys the Ricatti equation \cite{R1972} $\partial_t R /2 = k^* - k \, R^2$ solved with the initial condition $R(t_0)=0$. We have also defined the phase $\lambda = {\rm arg}\{\mu\}$ and introduced the phase $\tilde{\chi} = (\ii /2)\int_{t_0}^t dt' (\tb_0^* \partial_{t'} \tb_0 - \tb_0 \partial_{t'} \tb_0^*  )$, which is a generalization of the Berry phase found in a forced quantum harmonic oscillator \cite{CSS1987,paper357}.

At this point, we invoke the relation $\hat{\mathcal{S}}(t,t_0)=\ee^{\ii \hat{\mathcal{H}}_{\rm HO}^0 t/\hbar}\,\hat{\mathcal{U}}(t,t_0)\, \ee^{-\ii \hat{\mathcal{H}}_{\rm HO}^0t_0/\hbar}$, as well as $\ee^{\ii \hat{\mathcal{H}}_{\rm HO}^0 t_0/\hbar}\hat{D}\big(\tb_0 \ee^{\ii \omega_0 t_0 }\big)\ee^{-\ii \hat{\mathcal{H}}_{\rm HO}^0 t_0/\hbar}=\hat{D}\big(\tb_0 \big)$ and $\ee^{\ii \hat{\mathcal{H}}_{\rm HO}^0 t/\hbar}\hat{S}\left(\sigma_0 \ee^{ 2\ii \omega_0 t}\right)\ee^{-\ii \hat{\mathcal{H}}_{\rm HO}^0 t/\hbar}=\hat{S}\left(\sigma_0 \right)$, which can be easily verified by using ordering theorems for the displacement operator $\hat{D}\big( \tb_0\big)=\ee^{-|\tb_0|^2/2}\ee^{\tb_0^* a^\dagger}\ee^{-\tb_0 a}$ and the squeezing operator $\hat{S}(\sigma_0)$\,$=\exp\big[-\tanh(|\sigma_0|)\ee^{-\ii {\rm arg}\{\sigma_0\}}a^\dagger a^\dagger/2\big]$\,$\exp\{-\log \left[\cosh(|\sigma_0|)\right](a^\dagger a+a^\dagger a)\}$\,$\exp\big[\tanh(|\sigma_0|)\exp^{\ii {\rm arg}\{\sigma_0\}}a a/2\big]$, respectively \cite{BR97}. By employing these expressions, we find
\begin{align}
 \hat{\mathcal{S}}(t,t_0)=\ee^{\ii \tilde{\chi}} \, \ee^{\sigma_0\, a a /2 -\sigma_0^*a^\dagger a^\dagger/2 }\,\ee^{-\ii \lambda (a^\dagger a + a a^\dagger)/2}\,\ee^{\tb_0^* a^\dagger -\tb_0 a}. \label{St}
\end{align}
Finally, once the aforementioned transformation is applied to Eq.\ (\ref{St}) and the $t_0\rightarrow-\infty$ limit is taken, the coefficients $f_\ell^n$ in Eq.\ (\ref{fln}) of the main text are found by imposing the boundary condition $\psi_n(z \rightarrow -\infty,t)=\psi_0(\rb,t)\sum_{n=0}\alpha^0_n \ee^{-\ii n \omega_0 t}|n\rangle$, which is equivalent to $f_\ell^n(z\rightarrow -\infty)=\delta_{\ell,0}\alpha^0_n$.

\section{Phase-matched interaction}
\label{secB}
\renewcommand{\theequation}{B\arabic{equation}}

For an electron interacting with a structure that is translationally invariant along the e-beam propagation direction $z$, the electric field of the mode assumes the general functional form $\mathcal{E}_{0,i}(\rb)=\mathcal{E}_{0,i}(\Rb) \ee^{ \ii k_z z}$, with $\Rb=(x,y)$. We consider a guided polariton or a traveling wave of subluminal phase velocity in a high-refractive-index dielectric, such that the condition $\omega_0/v=k_z$ is met by tuning the electron velocity. This condition has been previously explored as a way to enhance the electron-mode coupling amplitude \cite{paper180}, which in the absence of ponderomotive interactions renders $\tb \propto \sqrt{L}$, where $L$ is the interaction length \cite{K19}.  In this scenario, we find the function $k= \ii \sum_i \eta_i^2$ with $\eta_i^2=(g_ie^2/\hbar \omega_0^2 \me v \gamma )\mathcal{E}_{0,i}^2(\Rb)$, which is independent of $z$ if we assume $\theta_0\sim 0$. This leads to the analytical solution $R=\ee^{- \ii {\rm arg}\{k\}}\tanh[2 |k|(z+L/2)]$ for the Ricatti equation in Appendix\ \ref{secA2}. By plugging $R$ into the coefficients $\mu$ and $\nu$, and then integrating over $z$ along the sample extension $L$, we obtain $\mu =\cosh[2|k|L]$ and $\nu = \ee^{-\ii \,{\rm arg}\{k\}}\sinh[2|k|L]$, from which the squeezing parameter reduces to Eq.\ (\ref{exactcoupl2}) in the main text (i.e., linear in $\eta^2_i$). To estimate $\tb_0$ up to the same order in $\eta_i^2$, we cannot disregard the phase $\theta_0$ in the calculation. By doing so, we obtain Eq.\ (\ref{exactcoupl1}), which approaches $\beta_0=(e L/\hbar \omega_0)\mathcal{E}_{0,z}(\Rb)$ for vanishing $\eta_i^2$.

\subsection{Hole in a metallic slab}
\label{secB1}

In order to estimate the squeezing coupling parameter $\sigma_0$ in a scenario of practical interest, we consider a circular hole of radius $a$ and length $L$ extending along the $z$ direction and drilled in a homogeneous metallic slab of permittivity $\epsilon(\omega)$, with the electron travelling in vacuum parallel to the hole axis. For $\omega_0 a/c\ll1$, we can neglect retardation in the description of the mode, which for an azimuthal number $m$, has an associated electric potential inside the hole ($R<a$) of the form $\phi_0(\rb)=A I_m(q_z R)\ee^{\ii q_zz+\ii m\varphi}$, and $\phi_0(\rb)=B K_m(q_z R)\ee^{\ii q_zz+\ii m\varphi}$ in the material region ($R>a$). These expressions are given in cylindrical coordinates $\rb=(R,\varphi,z)$ and involve the modified Bessel functions $K_m$ and $I_m$. Also, the constants $A$ and $B$ must satisfy the condition $B/A=I_m(q_za)/K_m(q_za)=\epsilon(\omega)I'_m(q_za)/K'_m(q_za)$ to guarantee the continuity of the potential and the normal electric displacement at the surface of the hole. We then calculate the corresponding electric field $\vec{\mathcal{E}}_0(\rb) = - \nabla \phi_0(\rb)$ and normalize the field as indicated in Sec.\ \ref{PMI}.

As an example, we take the electron to be focused at the hole center ($\Rb=0$, approaching with an azimuthal angle $\varphi$). Since excitations of symmetries $\pm m$ are degenerate, we can choose as a mode any linear combination of these two. In particular, we take $m=\pm1$ waves combined with a relative phase $\psi$, such that the mode field becomes $\vec{\mathcal{E}}^{m=\pm 1}_0(\Rb=0,z)=(-E_0/2^{3/2})\Big[\hat{\Rb}(1+\ee^{\ii(\psi -2 \varphi)})+\ii\, \hat{\boldsymbol{ \varphi}} (1- \ee^{\ii (\psi -2 \varphi)})\Big]\,\ee^{\ii q_z z +\ii \varphi }$, with $E_0=q_zA$ acting as a normalization constant. We estimate $E_0$ from the normalization condition $\int d^2\rb \,| \vec{\mathcal{E}}_0(\Rb)|^2=2\pi \hbar \omega_0/L$, neglecting the field inside the metal, which should be small because of the condition $|\epsilon(\omega_0)|\gg1$ required to have strong confinement. We obtain $E_0^2 =\hbar \omega_0/L a^2 I(\eta)$, where $\eta = \omega_0 a/v$ and $I(\eta) = \int_0^1 x\,dx\,\Big\{[I_0(x\eta)+I_2(x\eta)]^2/4+I_1^2(x\eta)(1/x^2\eta^2+1)\Big\}$. Plugging the field into the expressions for $k$ and $\sigma_0$ given above, we obtain the result shown at the end of Sec.\ \ref{PMI}.

\section{Proof of Eq.\ (9)}
\label{secC}
\renewcommand{\theequation}{C\arabic{equation}}

We start by considering the coefficient $F_\ell^n = \langle n| \,\hat{S}(\sigma_0)\, \ee^{-\lambda (a^\dagger a +a a^\dagger )/2}\, \hat{D}(\tb_0)\,|n + \ell \rangle $ under the following assumptions: ($i$) high bosonic population in the cavity, implying the condition $n+\ell \gg 1$; and ($ii$) small intrinsic electron-cavity couplings $|\sigma_0|\ll 1 $ and $|\tb_0|\ll 1$, yielding the condition $\ell/n \ll 1$. We use the ordering theorems mentioned in Appendix\ \ref{secA} together with the approximation ($ii$) to write $F_\ell^n \approx \ee^{-\ii \lambda /2 } \langle n| \ee^{- a^\dagger a^\dagger \sigma_0^*/2}\ee^{ a a \sigma_0/2}\ee^{-\ii \lambda a^\dagger a}\ee^{ \tb_0^* a^\dagger }\ee^{ -\tb_0 a}| n + \ell \rangle$.
Then, we expand the exponential operators in Taylor series to compute the corresponding matrix element, which leads to the result
\begin{align}
F_\ell^n \approx \,&\ee^{-\ii \lambda(n + \ell +\frac{1}{2} )}  \sum_{i,i',j =0}^\infty{\vphantom{\sum}}' \frac{(-1)^{ j+(3i' -\ell -i)/2}}{[j+(i' -\ell -i)/2]!j! i! i'!} \nonumber\\
&\times \left(\frac{|\sigma_0|}{2}\right)^{2j + (i' -\ell -i)/2} |\tb_0|^{i+i'} \nonumber\\
&\times \exp\big\{ \ii (i-i'+\ell ) \,{\rm arg}\{\sigma_0\}/2+ \ii (i'-i)({\rm arg}\{\tb_0\}+\lambda)\big\} \nonumber\\
&\times \frac{(n+\ell -i' +i)!\sqrt{(n+\ell)!n!}}{(n+\ell-i')!(n+\ell -i' +i -2 j)!}, \label{step1}
\end{align}
where the prime in the summation symbol indicates that the indices are restricted by the inequalities $n+\ell -i' +i-2j\geq 0$, $n+\ell -i' \geq 0$, and $j + (i'- i -\ell)/2\geq 0$, as well as by the condition that $i'-\ell -i$ is an even number. In virtue of assumption ($ii$), the first two inequalities are automatically satisfied. Also, this assumption allows us to follow a procedure similar to the one described in the Methods section of Ref.\ \cite{paper360}, consisting in using the Striling formula to approximate the factorials and neglecting all the indices in front of $n$ when they are not exponentiated. Then, the entire factor in the fourth line (exponential factor) of Eq.\ (\ref{step1}) reduces to $(n+\ell)^{i'+2j-\ell/2}$, and by making the substitutions $m=(i'-i-\ell)/2$ and $s=j+m$, we obtain 
\begin{align}
F_\ell^n \approx \,& \ee^{\ii \ell {\rm arg}\{-\tb_0\}-\ii \lambda (m+\frac{1}{2})} \sum_{i=0}^\infty{\vphantom{\sum}}'  \sum\limits_{\substack{m=-(i+\ell)/2\\s=\max\{m,0\}}}^\infty (-1)^{i+s} \label{step2}\\
&\times \left(\frac{|\sigma_0|}{2}\right)^{2s-m}|\tb_0|^{2(i+m)+\ell} \frac{(n+\ell)^{2s+i+\ell/2}}{s!(s-m)! i! (2m+i+\ell)!}, \nonumber
\end{align}
where now the prime indicates that the leftmost sum is restricted to even values of $i+\ell$. Finally, by pushing the lower limits of the $m$ and $s$ sums to $m=-\infty$ and $s=0$, and further using the series expansion of the Bessel functions \cite{GR1980} $J_\ell(x)=\sum_{m=0}^\infty (-1)^m (x/2)^{2m+\ell}/(m+\ell)!m!$, we find that Eq.\ (\ref{step2}) directly reduces to Eq.\ (\ref{Pnextpump}).

\section*{Acknowledgments}
This work has been supported in part by the European Research Council (Advanced Grant 789104-eNANO), European Commission (Horizon 2020 Grants 101017720 FET-Proactive EBEAM and 964591-SMART-electron), Spanish MICINN (PID2020-112625GB-I00 and Severo Ochoa CEX2019-000910-S), Catalan CERCA Program, and Fundaci\'{o}s Cellex and Mir-Puig.

\begin{thebibliography}{48}
\expandafter\ifx\csname natexlab\endcsname\relax\def\natexlab#1{#1}\fi
\expandafter\ifx\csname bibnamefont\endcsname\relax
  \def\bibnamefont#1{#1}\fi
\expandafter\ifx\csname bibfnamefont\endcsname\relax
  \def\bibfnamefont#1{#1}\fi
\expandafter\ifx\csname citenamefont\endcsname\relax
  \def\citenamefont#1{#1}\fi
\expandafter\ifx\csname url\endcsname\relax
  \def\url#1{\texttt{#1}}\fi
\expandafter\ifx\csname urlprefix\endcsname\relax\def\urlprefix{URL }\fi
\providecommand{\bibinfo}[2]{#2}
\providecommand{\eprint}[2][]{\url{#2}}

\bibitem[{\citenamefont{Walls}(1983)}]{W1983}
\bibinfo{author}{\bibfnamefont{D.~F.} \bibnamefont{Walls}},
  \bibinfo{journal}{Nature} \textbf{\bibinfo{volume}{306}},
  \bibinfo{pages}{141} (\bibinfo{year}{1983}).

\bibitem[{\citenamefont{Gottesman et~al.}(2001)\citenamefont{Gottesman, Kitaev,
  and Preskill}}]{GKP01}
\bibinfo{author}{\bibfnamefont{D.}~\bibnamefont{Gottesman}},
  \bibinfo{author}{\bibfnamefont{A.}~\bibnamefont{Kitaev}}, \bibnamefont{and}
  \bibinfo{author}{\bibfnamefont{J.}~\bibnamefont{Preskill}},
  \bibinfo{journal}{Phys.\ Rev.\ A} \textbf{\bibinfo{volume}{64}},
  \bibinfo{pages}{012310} (\bibinfo{year}{2001}).

\bibitem[{\citenamefont{Menicucci}(2014)}]{M14_3}
\bibinfo{author}{\bibfnamefont{N.~C.} \bibnamefont{Menicucci}},
  \bibinfo{journal}{Phys.\ Rev.\ Lett.} \textbf{\bibinfo{volume}{112}},
  \bibinfo{pages}{120504} (\bibinfo{year}{2014}).

\bibitem[{\citenamefont{Slusher et~al.}(1985)\citenamefont{Slusher, Hollberg,
  Yurke, Mertz, and Valley}}]{SHY1985}
\bibinfo{author}{\bibfnamefont{R.~E.} \bibnamefont{Slusher}},
  \bibinfo{author}{\bibfnamefont{L.~W.} \bibnamefont{Hollberg}},
  \bibinfo{author}{\bibfnamefont{B.}~\bibnamefont{Yurke}},
  \bibinfo{author}{\bibfnamefont{J.~C.} \bibnamefont{Mertz}}, \bibnamefont{and}
  \bibinfo{author}{\bibfnamefont{J.~F.} \bibnamefont{Valley}},
  \bibinfo{journal}{Phys.\ Rev.\ Lett.} \textbf{\bibinfo{volume}{55}},
  \bibinfo{pages}{2409} (\bibinfo{year}{1985}).

\bibitem[{\citenamefont{Wu et~al.}(1986)\citenamefont{Wu, Kimble, Hall, and
  Wu}}]{WKH1986}
\bibinfo{author}{\bibfnamefont{L.-A.} \bibnamefont{Wu}},
  \bibinfo{author}{\bibfnamefont{H.~J.} \bibnamefont{Kimble}},
  \bibinfo{author}{\bibfnamefont{J.~L.} \bibnamefont{Hall}}, \bibnamefont{and}
  \bibinfo{author}{\bibfnamefont{H.}~\bibnamefont{Wu}},
  \bibinfo{journal}{Phys.\ Rev.\ Lett.} \textbf{\bibinfo{volume}{57}},
  \bibinfo{pages}{2520} (\bibinfo{year}{1986}).

\bibitem[{\citenamefont{{Benda\~{n}a} et~al.}(2011)\citenamefont{{Benda\~{n}a},
  Polman, and {Garc\'{\i}a de Abajo}}}]{paper180}
\bibinfo{author}{\bibfnamefont{X.~M.} \bibnamefont{{Benda\~{n}a}}},
  \bibinfo{author}{\bibfnamefont{A.}~\bibnamefont{Polman}}, \bibnamefont{and}
  \bibinfo{author}{\bibfnamefont{F.~J.} \bibnamefont{{Garc\'{\i}a de Abajo}}},
  \bibinfo{journal}{Nano\ Lett.} \textbf{\bibinfo{volume}{11}},
  \bibinfo{pages}{5099} (\bibinfo{year}{2011}).

\bibitem[{\citenamefont{Feist et~al.}(2022)\citenamefont{Feist, Huang, Arend,
  Yang, Henke, Raja, Kappert, Wang, {Louren{\c{c}}o-Martins}, Qiu
  et~al.}}]{FHA22}
\bibinfo{author}{\bibfnamefont{A.}~\bibnamefont{Feist}},
  \bibinfo{author}{\bibfnamefont{G.}~\bibnamefont{Huang}},
  \bibinfo{author}{\bibfnamefont{G.}~\bibnamefont{Arend}},
  \bibinfo{author}{\bibfnamefont{Y.}~\bibnamefont{Yang}},
  \bibinfo{author}{\bibfnamefont{J.-W.} \bibnamefont{Henke}},
  \bibinfo{author}{\bibfnamefont{A.~S.} \bibnamefont{Raja}},
  \bibinfo{author}{\bibfnamefont{F.~J.} \bibnamefont{Kappert}},
  \bibinfo{author}{\bibfnamefont{R.~N.} \bibnamefont{Wang}},
  \bibinfo{author}{\bibfnamefont{H.}~\bibnamefont{{Louren{\c{c}}o-Martins}}},
  \bibinfo{author}{\bibfnamefont{Z.}~\bibnamefont{Qiu}}, \bibnamefont{et~al.},
  \emph{\bibinfo{title}{Cavity-mediated electron-photon pairs}}
  (\bibinfo{year}{2022}), \eprint{2202.12821}.

\bibitem[{\citenamefont{{Garc\'{\i}a de Abajo}}(2010)}]{paper149}
\bibinfo{author}{\bibfnamefont{F.~J.} \bibnamefont{{Garc\'{\i}a de Abajo}}},
  \bibinfo{journal}{Rev.\ Mod.\ Phys.} \textbf{\bibinfo{volume}{82}},
  \bibinfo{pages}{209} (\bibinfo{year}{2010}).

\bibitem[{\citenamefont{Becker et~al.}(1982)\citenamefont{Becker, Scully, and
  Zubairy}}]{BSZ1982}
\bibinfo{author}{\bibfnamefont{W.}~\bibnamefont{Becker}},
  \bibinfo{author}{\bibfnamefont{M.~O.} \bibnamefont{Scully}},
  \bibnamefont{and} \bibinfo{author}{\bibfnamefont{M.~S.}
  \bibnamefont{Zubairy}}, \bibinfo{journal}{Phys.\ Rev.\ Lett.}
  \textbf{\bibinfo{volume}{48}}, \bibinfo{pages}{475} (\bibinfo{year}{1982}).

\bibitem[{\citenamefont{Gjaja and Bhattacharjee}(1987)}]{GB1987}
\bibinfo{author}{\bibfnamefont{I.}~\bibnamefont{Gjaja}} \bibnamefont{and}
  \bibinfo{author}{\bibfnamefont{A.}~\bibnamefont{Bhattacharjee}},
  \bibinfo{journal}{Phys.\ Rev.\ A} \textbf{\bibinfo{volume}{36}},
  \bibinfo{pages}{5486} (\bibinfo{year}{1987}).

\bibitem[{\citenamefont{Gjaja and Bhattacharjee}(1991)}]{GB91}
\bibinfo{author}{\bibfnamefont{I.}~\bibnamefont{Gjaja}} \bibnamefont{and}
  \bibinfo{author}{\bibfnamefont{A.}~\bibnamefont{Bhattacharjee}},
  \bibinfo{journal}{Phys.\ Rev.\ A} \textbf{\bibinfo{volume}{43}},
  \bibinfo{pages}{3206} (\bibinfo{year}{1991}).

\bibitem[{\citenamefont{Lobastov et~al.}(2005)\citenamefont{Lobastov,
  Srinivasan, and Zewail}}]{LSZ05}
\bibinfo{author}{\bibfnamefont{V.~A.} \bibnamefont{Lobastov}},
  \bibinfo{author}{\bibfnamefont{R.}~\bibnamefont{Srinivasan}},
  \bibnamefont{and} \bibinfo{author}{\bibfnamefont{A.~H.}
  \bibnamefont{Zewail}}, \bibinfo{journal}{Proc.\ Natl.\ Academ.\ Sci.}
  \textbf{\bibinfo{volume}{102}}, \bibinfo{pages}{7069} (\bibinfo{year}{2005}).

\bibitem[{\citenamefont{Barwick and Zewail}(2015)}]{BZ15}
\bibinfo{author}{\bibfnamefont{B.}~\bibnamefont{Barwick}} \bibnamefont{and}
  \bibinfo{author}{\bibfnamefont{A.~H.} \bibnamefont{Zewail}},
  \bibinfo{journal}{ACS\ Photonics} \textbf{\bibinfo{volume}{2}},
  \bibinfo{pages}{1391} (\bibinfo{year}{2015}).

\bibitem[{\citenamefont{Aseyev et~al.}(2020)\citenamefont{Aseyev, Ryabov,
  Mironov, and Ischenko}}]{ARM20}
\bibinfo{author}{\bibfnamefont{S.~A.} \bibnamefont{Aseyev}},
  \bibinfo{author}{\bibfnamefont{E.~A.} \bibnamefont{Ryabov}},
  \bibinfo{author}{\bibfnamefont{B.~N.} \bibnamefont{Mironov}},
  \bibnamefont{and} \bibinfo{author}{\bibfnamefont{A.~A.}
  \bibnamefont{Ischenko}}, \bibinfo{journal}{Crystals}
  \textbf{\bibinfo{volume}{10}}, \bibinfo{pages}{452} (\bibinfo{year}{2020}).

\bibitem[{\citenamefont{Barwick et~al.}(2009)\citenamefont{Barwick, Flannigan,
  and Zewail}}]{BFZ09}
\bibinfo{author}{\bibfnamefont{B.}~\bibnamefont{Barwick}},
  \bibinfo{author}{\bibfnamefont{D.~J.} \bibnamefont{Flannigan}},
  \bibnamefont{and} \bibinfo{author}{\bibfnamefont{A.~H.}
  \bibnamefont{Zewail}}, \bibinfo{journal}{Nature}
  \textbf{\bibinfo{volume}{462}}, \bibinfo{pages}{902} (\bibinfo{year}{2009}).

\bibitem[{\citenamefont{Feist et~al.}(2015)\citenamefont{Feist, Echternkamp,
  Schauss, Yalunin, Sch\"afer, and Ropers}}]{FES15}
\bibinfo{author}{\bibfnamefont{A.}~\bibnamefont{Feist}},
  \bibinfo{author}{\bibfnamefont{K.~E.} \bibnamefont{Echternkamp}},
  \bibinfo{author}{\bibfnamefont{J.}~\bibnamefont{Schauss}},
  \bibinfo{author}{\bibfnamefont{S.~V.} \bibnamefont{Yalunin}},
  \bibinfo{author}{\bibfnamefont{S.}~\bibnamefont{Sch\"afer}},
  \bibnamefont{and} \bibinfo{author}{\bibfnamefont{C.}~\bibnamefont{Ropers}},
  \bibinfo{journal}{Nature} \textbf{\bibinfo{volume}{521}},
  \bibinfo{pages}{200} (\bibinfo{year}{2015}).

\bibitem[{\citenamefont{Priebe et~al.}(2017)\citenamefont{Priebe, Rathje,
  Yalunin, Hohage, Feist, Sch\"{a}fer, and Ropers}}]{PRY17}
\bibinfo{author}{\bibfnamefont{K.~E.} \bibnamefont{Priebe}},
  \bibinfo{author}{\bibfnamefont{C.}~\bibnamefont{Rathje}},
  \bibinfo{author}{\bibfnamefont{S.~V.} \bibnamefont{Yalunin}},
  \bibinfo{author}{\bibfnamefont{T.}~\bibnamefont{Hohage}},
  \bibinfo{author}{\bibfnamefont{A.}~\bibnamefont{Feist}},
  \bibinfo{author}{\bibfnamefont{S.}~\bibnamefont{Sch\"{a}fer}},
  \bibnamefont{and} \bibinfo{author}{\bibfnamefont{C.}~\bibnamefont{Ropers}},
  \bibinfo{journal}{Nat.\ Photon.} \textbf{\bibinfo{volume}{11}},
  \bibinfo{pages}{793} (\bibinfo{year}{2017}).

\bibitem[{\citenamefont{Morimoto and Baum}(2018)}]{MB18_2}
\bibinfo{author}{\bibfnamefont{Y.}~\bibnamefont{Morimoto}} \bibnamefont{and}
  \bibinfo{author}{\bibfnamefont{P.}~\bibnamefont{Baum}},
  \bibinfo{journal}{Nature\ Phys.} \textbf{\bibinfo{volume}{14}},
  \bibinfo{pages}{252} (\bibinfo{year}{2018}).

\bibitem[{\citenamefont{Baum}(2017)}]{B17_2}
\bibinfo{author}{\bibfnamefont{P.}~\bibnamefont{Baum}}, \bibinfo{journal}{J.\
  Appl.\ Phys.} \textbf{\bibinfo{volume}{122}}, \bibinfo{pages}{223105}
  (\bibinfo{year}{2017}).

\bibitem[{\citenamefont{{Di Giulio} and {Garc\'{\i}a de
  Abajo}}(2020{\natexlab{a}})}]{paper360}
\bibinfo{author}{\bibfnamefont{V.}~\bibnamefont{{Di Giulio}}} \bibnamefont{and}
  \bibinfo{author}{\bibfnamefont{F.~J.} \bibnamefont{{Garc\'{\i}a de Abajo}}},
  \bibinfo{journal}{Optica} \textbf{\bibinfo{volume}{7}}, \bibinfo{pages}{1820}
  (\bibinfo{year}{2020}{\natexlab{a}}).

\bibitem[{\citenamefont{{Di Giulio} et~al.}(2019)\citenamefont{{Di Giulio},
  Kociak, and {Garc\'{\i}a de Abajo}}}]{paper339}
\bibinfo{author}{\bibfnamefont{V.}~\bibnamefont{{Di Giulio}}},
  \bibinfo{author}{\bibfnamefont{M.}~\bibnamefont{Kociak}}, \bibnamefont{and}
  \bibinfo{author}{\bibfnamefont{F.~J.} \bibnamefont{{Garc\'{\i}a de Abajo}}},
  \bibinfo{journal}{Optica} \textbf{\bibinfo{volume}{6}}, \bibinfo{pages}{1524}
  (\bibinfo{year}{2019}).

\bibitem[{\citenamefont{Hayun et~al.}(2021)\citenamefont{Hayun, Reinhardt,
  Nemirovsky, Karnieli, Rivera, and Kaminer}}]{HRN21}
\bibinfo{author}{\bibfnamefont{A.~B.} \bibnamefont{Hayun}},
  \bibinfo{author}{\bibfnamefont{O.}~\bibnamefont{Reinhardt}},
  \bibinfo{author}{\bibfnamefont{J.}~\bibnamefont{Nemirovsky}},
  \bibinfo{author}{\bibfnamefont{A.}~\bibnamefont{Karnieli}},
  \bibinfo{author}{\bibfnamefont{N.}~\bibnamefont{Rivera}}, \bibnamefont{and}
  \bibinfo{author}{\bibfnamefont{I.}~\bibnamefont{Kaminer}},
  \bibinfo{journal}{Sci.\ Adv.} \textbf{\bibinfo{volume}{7}},
  \bibinfo{pages}{eabe4270} (\bibinfo{year}{2021}).

\bibitem[{\citenamefont{Dahan et~al.}(2022)\citenamefont{Dahan, Baranes, Ruimy,
  Rivera, and .Kaminer}}]{DBG22}
\bibinfo{author}{\bibfnamefont{R.}~\bibnamefont{Dahan}},
  \bibinfo{author}{\bibfnamefont{G.}~\bibnamefont{Baranes}},
  \bibinfo{author}{\bibfnamefont{A.~G.~R.} \bibnamefont{Ruimy}},
  \bibinfo{author}{\bibfnamefont{N.}~\bibnamefont{Rivera}}, \bibnamefont{and}
  \bibinfo{author}{\bibfnamefont{I.}~\bibnamefont{.Kaminer}},
  \bibinfo{journal}{arXiv} p. \bibinfo{pages}{2206.08828}
  (\bibinfo{year}{2022}).

\bibitem[{\citenamefont{Rocca}(1995)}]{R95}
\bibinfo{author}{\bibfnamefont{M.}~\bibnamefont{Rocca}},
  \bibinfo{journal}{Surf.\ Sci.\ Rep.} \textbf{\bibinfo{volume}{22}},
  \bibinfo{pages}{1} (\bibinfo{year}{1995}).

\bibitem[{\citenamefont{Nagao et~al.}(2006)\citenamefont{Nagao, Yaginuma,
  Inaoka, and Sakurai}}]{NYI06}
\bibinfo{author}{\bibfnamefont{T.}~\bibnamefont{Nagao}},
  \bibinfo{author}{\bibfnamefont{S.}~\bibnamefont{Yaginuma}},
  \bibinfo{author}{\bibfnamefont{T.}~\bibnamefont{Inaoka}}, \bibnamefont{and}
  \bibinfo{author}{\bibfnamefont{T.}~\bibnamefont{Sakurai}},
  \bibinfo{journal}{Phys.\ Rev.\ Lett.} \textbf{\bibinfo{volume}{97}},
  \bibinfo{pages}{116802} (\bibinfo{year}{2006}).

\bibitem[{\citenamefont{Yuen}(1976)}]{Y1976}
\bibinfo{author}{\bibfnamefont{H.~P.} \bibnamefont{Yuen}},
  \bibinfo{journal}{Phys.\ Rev.\ A} \textbf{\bibinfo{volume}{13}},
  \bibinfo{pages}{2226} (\bibinfo{year}{1976}).

\bibitem[{\citenamefont{{Di Giulio} and {Garc\'{\i}a de
  Abajo}}(2020{\natexlab{b}})}]{paper357}
\bibinfo{author}{\bibfnamefont{V.}~\bibnamefont{{Di Giulio}}} \bibnamefont{and}
  \bibinfo{author}{\bibfnamefont{F.~J.} \bibnamefont{{Garc\'{\i}a de Abajo}}},
  \bibinfo{journal}{New\ J.\ Phys.} \textbf{\bibinfo{volume}{22}},
  \bibinfo{pages}{103057} (\bibinfo{year}{2020}{\natexlab{b}}).

\bibitem[{\citenamefont{Reid}(1972)}]{R1972}
\bibinfo{author}{\bibfnamefont{W.~T.} \bibnamefont{Reid}},
  \emph{\bibinfo{title}{Ricatti Differential Equations}}
  (\bibinfo{publisher}{Academic Press}, \bibinfo{address}{111 5th Avenue, New
  York, New York}, \bibinfo{year}{1972}).

\bibitem[{\citenamefont{Bishop and Vourdas}(1986)}]{BV1986}
\bibinfo{author}{\bibfnamefont{R.~F.} \bibnamefont{Bishop}} \bibnamefont{and}
  \bibinfo{author}{\bibfnamefont{A.}~\bibnamefont{Vourdas}},
  \bibinfo{journal}{J.\ Phys.\ A:\ Math.\ Gen.} \textbf{\bibinfo{volume}{19}},
  \bibinfo{pages}{2525} (\bibinfo{year}{1986}).

\bibitem[{\citenamefont{Glauber and Lewenstein}(1991)}]{GL91}
\bibinfo{author}{\bibfnamefont{R.~J.} \bibnamefont{Glauber}} \bibnamefont{and}
  \bibinfo{author}{\bibfnamefont{M.}~\bibnamefont{Lewenstein}},
  \bibinfo{journal}{Phys.\ Rev.\ A} \textbf{\bibinfo{volume}{43}},
  \bibinfo{pages}{467} (\bibinfo{year}{1991}).

\bibitem[{\citenamefont{Kfir}(2019)}]{K19}
\bibinfo{author}{\bibfnamefont{O.}~\bibnamefont{Kfir}},
  \bibinfo{journal}{Phys.\ Rev.\ Lett.} \textbf{\bibinfo{volume}{123}},
  \bibinfo{pages}{103602} (\bibinfo{year}{2019}).

\bibitem[{\citenamefont{{Di Giulio} et~al.}(2021)\citenamefont{{Di Giulio},
  Kfir, Ropers, and {Garc\'{\i}a de Abajo}}}]{paper373}
\bibinfo{author}{\bibfnamefont{V.}~\bibnamefont{{Di Giulio}}},
  \bibinfo{author}{\bibfnamefont{O.}~\bibnamefont{Kfir}},
  \bibinfo{author}{\bibfnamefont{C.}~\bibnamefont{Ropers}}, \bibnamefont{and}
  \bibinfo{author}{\bibfnamefont{F.~J.} \bibnamefont{{Garc\'{\i}a de Abajo}}},
  \bibinfo{journal}{ACS\ Nano} \textbf{\bibinfo{volume}{15}},
  \bibinfo{pages}{7290} (\bibinfo{year}{2021}).

\bibitem[{\citenamefont{Kfir et~al.}(2021)\citenamefont{Kfir, {Di Giulio},
  {Garc\'{\i}a de Abajo}, and Ropers}}]{paper374}
\bibinfo{author}{\bibfnamefont{O.}~\bibnamefont{Kfir}},
  \bibinfo{author}{\bibfnamefont{V.}~\bibnamefont{{Di Giulio}}},
  \bibinfo{author}{\bibfnamefont{F.~J.} \bibnamefont{{Garc\'{\i}a de Abajo}}},
  \bibnamefont{and} \bibinfo{author}{\bibfnamefont{C.}~\bibnamefont{Ropers}},
  \bibinfo{journal}{Sci.\ Adv.} \textbf{\bibinfo{volume}{7}},
  \bibinfo{pages}{eabf6380} (\bibinfo{year}{2021}).

\bibitem[{\citenamefont{Nielsen and Chuang}(2004)}]{NC04}
\bibinfo{author}{\bibfnamefont{M.~A.} \bibnamefont{Nielsen}} \bibnamefont{and}
  \bibinfo{author}{\bibfnamefont{I.~L.} \bibnamefont{Chuang}},
  \emph{\bibinfo{title}{Quantum Computation and Quantum Information (Cambridge
  Series on Information and the Natural Sciences)}}
  (\bibinfo{publisher}{Cambridge University Press}, \bibinfo{year}{2004}),
  \bibinfo{edition}{1st} ed., ISBN \bibinfo{isbn}{0521635039}.

\bibitem[{\citenamefont{Zhao et~al.}(2021)\citenamefont{Zhao, Sun, and
  Fan}}]{ZSF21}
\bibinfo{author}{\bibfnamefont{Z.}~\bibnamefont{Zhao}},
  \bibinfo{author}{\bibfnamefont{X.-Q.} \bibnamefont{Sun}}, \bibnamefont{and}
  \bibinfo{author}{\bibfnamefont{S.}~\bibnamefont{Fan}},
  \emph{\bibinfo{title}{Quantum entanglement and modulation enhancement of
  free-electron--bound-electron interaction}} (\bibinfo{year}{2021}).

\bibitem[{\citenamefont{Yalunin et~al.}(2021)\citenamefont{Yalunin, Feist, and
  Ropers}}]{YFR21}
\bibinfo{author}{\bibfnamefont{S.~V.} \bibnamefont{Yalunin}},
  \bibinfo{author}{\bibfnamefont{A.}~\bibnamefont{Feist}}, \bibnamefont{and}
  \bibinfo{author}{\bibfnamefont{C.}~\bibnamefont{Ropers}},
  \bibinfo{journal}{Phys.\ Rev.\ Research} \textbf{\bibinfo{volume}{3}},
  \bibinfo{pages}{L032036} (\bibinfo{year}{2021}).

\bibitem[{\citenamefont{Bourassa et~al.}(2020)\citenamefont{Bourassa,
  Menicucci, and Sabapathy}}]{TBM20}
\bibinfo{author}{\bibfnamefont{I.~T. J.~E.} \bibnamefont{Bourassa}},
  \bibinfo{author}{\bibfnamefont{N.~C.} \bibnamefont{Menicucci}},
  \bibnamefont{and} \bibinfo{author}{\bibfnamefont{K.~K.}
  \bibnamefont{Sabapathy}}, \bibinfo{journal}{Phys.\ Rev.\ A}
  \textbf{\bibinfo{volume}{101}}, \bibinfo{pages}{032315}
  (\bibinfo{year}{2020}).

\bibitem[{\citenamefont{Echternkamp et~al.}(2016)\citenamefont{Echternkamp,
  Feist, Sch\"{a}fer, and Ropers}}]{EFS16}
\bibinfo{author}{\bibfnamefont{K.~E.} \bibnamefont{Echternkamp}},
  \bibinfo{author}{\bibfnamefont{A.}~\bibnamefont{Feist}},
  \bibinfo{author}{\bibfnamefont{S.}~\bibnamefont{Sch\"{a}fer}},
  \bibnamefont{and} \bibinfo{author}{\bibfnamefont{C.}~\bibnamefont{Ropers}},
  \bibinfo{journal}{Nat.\ Phys.} \textbf{\bibinfo{volume}{12}},
  \bibinfo{pages}{1000} (\bibinfo{year}{2016}).

\bibitem[{\citenamefont{Kone\v{c}n\'{a}
  et~al.}(2019)\citenamefont{Kone\v{c}n\'{a}, {Di Giulio}, Mkhitaryan, Ropers,
  and {Garc\'{\i}a de Abajo}}}]{paper347}
\bibinfo{author}{\bibfnamefont{A.}~\bibnamefont{Kone\v{c}n\'{a}}},
  \bibinfo{author}{\bibfnamefont{V.}~\bibnamefont{{Di Giulio}}},
  \bibinfo{author}{\bibfnamefont{V.}~\bibnamefont{Mkhitaryan}},
  \bibinfo{author}{\bibfnamefont{C.}~\bibnamefont{Ropers}}, \bibnamefont{and}
  \bibinfo{author}{\bibfnamefont{F.~J.} \bibnamefont{{Garc\'{\i}a de Abajo}}},
  \bibinfo{journal}{ACS\ Photonics} \textbf{\bibinfo{volume}{7}},
  \bibinfo{pages}{1290} (\bibinfo{year}{2019}).

\bibitem[{\citenamefont{Kfir et~al.}(2020)\citenamefont{Kfir,
  Louren\c{c}o-Martins, Storeck, Sivis, Harvey, Kippenberg, Feist, and
  Ropers}}]{KLS20}
\bibinfo{author}{\bibfnamefont{O.}~\bibnamefont{Kfir}},
  \bibinfo{author}{\bibfnamefont{H.}~\bibnamefont{Louren\c{c}o-Martins}},
  \bibinfo{author}{\bibfnamefont{G.}~\bibnamefont{Storeck}},
  \bibinfo{author}{\bibfnamefont{M.}~\bibnamefont{Sivis}},
  \bibinfo{author}{\bibfnamefont{T.~R.} \bibnamefont{Harvey}},
  \bibinfo{author}{\bibfnamefont{T.~J.} \bibnamefont{Kippenberg}},
  \bibinfo{author}{\bibfnamefont{A.}~\bibnamefont{Feist}}, \bibnamefont{and}
  \bibinfo{author}{\bibfnamefont{C.}~\bibnamefont{Ropers}},
  \bibinfo{journal}{Nature} \textbf{\bibinfo{volume}{582}}, \bibinfo{pages}{46}
  (\bibinfo{year}{2020}).

\bibitem[{\citenamefont{Dahan et~al.}(2020)\citenamefont{Dahan, Nehemia,
  Shentcis, Reinhardt, Adiv, Shi, Be'er, Lynch, Kurman, Wang et~al.}}]{DNS20}
\bibinfo{author}{\bibfnamefont{R.}~\bibnamefont{Dahan}},
  \bibinfo{author}{\bibfnamefont{S.}~\bibnamefont{Nehemia}},
  \bibinfo{author}{\bibfnamefont{M.}~\bibnamefont{Shentcis}},
  \bibinfo{author}{\bibfnamefont{O.}~\bibnamefont{Reinhardt}},
  \bibinfo{author}{\bibfnamefont{Y.}~\bibnamefont{Adiv}},
  \bibinfo{author}{\bibfnamefont{X.}~\bibnamefont{Shi}},
  \bibinfo{author}{\bibfnamefont{O.}~\bibnamefont{Be'er}},
  \bibinfo{author}{\bibfnamefont{M.~H.} \bibnamefont{Lynch}},
  \bibinfo{author}{\bibfnamefont{Y.}~\bibnamefont{Kurman}},
  \bibinfo{author}{\bibfnamefont{K.}~\bibnamefont{Wang}}, \bibnamefont{et~al.},
  \bibinfo{journal}{Nature\ Phys.} \textbf{\bibinfo{volume}{16}},
  \bibinfo{pages}{1123} (\bibinfo{year}{2020}).

\bibitem[{\citenamefont{M\"{u}ller et~al.}(2021)\citenamefont{M\"{u}ller, Hock,
  Koch, Bach, Rathje, and Sch\"{a}fer}}]{MHK21}
\bibinfo{author}{\bibfnamefont{N.}~\bibnamefont{M\"{u}ller}},
  \bibinfo{author}{\bibfnamefont{V.}~\bibnamefont{Hock}},
  \bibinfo{author}{\bibfnamefont{H.}~\bibnamefont{Koch}},
  \bibinfo{author}{\bibfnamefont{N.}~\bibnamefont{Bach}},
  \bibinfo{author}{\bibfnamefont{C.}~\bibnamefont{Rathje}}, \bibnamefont{and}
  \bibinfo{author}{\bibfnamefont{S.}~\bibnamefont{Sch\"{a}fer}},
  \bibinfo{journal}{ACS\ Photonics} \textbf{\bibinfo{volume}{8}},
  \bibinfo{pages}{1569} (\bibinfo{year}{2021}).

\bibitem[{\citenamefont{Huang et~al.}(2022)\citenamefont{Huang, Engelsen, Kfir,
  Ropers, and Kippenberg}}]{HEK22}
\bibinfo{author}{\bibfnamefont{G.}~\bibnamefont{Huang}},
  \bibinfo{author}{\bibfnamefont{N.~J.} \bibnamefont{Engelsen}},
  \bibinfo{author}{\bibfnamefont{O.}~\bibnamefont{Kfir}},
  \bibinfo{author}{\bibfnamefont{C.}~\bibnamefont{Ropers}}, \bibnamefont{and}
  \bibinfo{author}{\bibfnamefont{T.~J.} \bibnamefont{Kippenberg}},
  \bibinfo{journal}{arXiv} p. \bibinfo{pages}{2206.08098}
  (\bibinfo{year}{2022}).

\bibitem[{\citenamefont{{Garc\'{\i}a de Abajo} and
  Kone\v{c}n\'{a}}(2021)}]{paper368}
\bibinfo{author}{\bibfnamefont{F.~J.} \bibnamefont{{Garc\'{\i}a de Abajo}}}
  \bibnamefont{and}
  \bibinfo{author}{\bibfnamefont{A.}~\bibnamefont{Kone\v{c}n\'{a}}},
  \bibinfo{journal}{Phys.\ Rev.\ Lett.} \textbf{\bibinfo{volume}{126}},
  \bibinfo{pages}{123901} (\bibinfo{year}{2021}).

\bibitem[{\citenamefont{Sakurai}(1994)}]{S94}
\bibinfo{author}{\bibfnamefont{J.~J.} \bibnamefont{Sakurai}},
  \emph{\bibinfo{title}{Modern Quantum Mechanics}}
  (\bibinfo{publisher}{Addison-Wesley}, \bibinfo{year}{1994}).

\bibitem[{\citenamefont{Chaturvedi et~al.}(1987)\citenamefont{Chaturvedi,
  Sriram, and Srinivasan}}]{CSS1987}
\bibinfo{author}{\bibfnamefont{S.}~\bibnamefont{Chaturvedi}},
  \bibinfo{author}{\bibfnamefont{M.~S.} \bibnamefont{Sriram}},
  \bibnamefont{and}
  \bibinfo{author}{\bibfnamefont{V.}~\bibnamefont{Srinivasan}},
  \bibinfo{journal}{J.\ Phys.\ A:\ Math.\ Gen.} \textbf{\bibinfo{volume}{20}},
  \bibinfo{pages}{L1071} (\bibinfo{year}{1987}).

\bibitem[{\citenamefont{Barnett and Radmore}(1997)}]{BR97}
\bibinfo{author}{\bibfnamefont{S.~M.} \bibnamefont{Barnett}} \bibnamefont{and}
  \bibinfo{author}{\bibfnamefont{P.~M.} \bibnamefont{Radmore}},
  \emph{\bibinfo{title}{Methods in Theoretical Quantum Optics}}
  (\bibinfo{publisher}{Oxford University Press}, \bibinfo{address}{Oxford},
  \bibinfo{year}{1997}).

\bibitem[{\citenamefont{Gradshteyn and Ryzhik}(2007)}]{GR1980}
\bibinfo{author}{\bibfnamefont{I.~S.} \bibnamefont{Gradshteyn}}
  \bibnamefont{and} \bibinfo{author}{\bibfnamefont{I.~M.}
  \bibnamefont{Ryzhik}}, \emph{\bibinfo{title}{Table of Integrals, Series, and
  Products}} (\bibinfo{publisher}{Academic Press}, \bibinfo{address}{London},
  \bibinfo{year}{2007}).

\end{thebibliography}

\end{document}